\documentclass[11pt]{article}
\usepackage{jheppub}
\usepackage{amsmath,amssymb,amsthm,mathtools,booktabs,tabularray,graphicx,mathrsfs,comment}
\usepackage{tikz}
\usetikzlibrary{calc}
\definecolor{tubeRed}{RGB}{241,101,89}
\definecolor{tubeBlue}{RGB}{105,158,242}
\definecolor{tubeGreen}{RGB}{126,185,76}
\usepackage{xcolor,microtype}
\usepackage{environ}

\UseTblrLibrary{booktabs}
\hypersetup{colorlinks=true,linkcolor=blue,citecolor=blue,urlcolor=blue}
\allowdisplaybreaks
\newcommand{\dd}{\mathrm d}

\newcommand{\adj}{\operatorname{adj}}

\newcommand{\Disc}{\operatorname{Disc}}

\NewEnviron{eqn}{
\begin{align}
\begin{split}
  \BODY
\end{split}
\end{align}
}

\NewEnviron{eqn*}{
\begin{align*}
\begin{split}
  \BODY
\end{split}
\end{align*}
}

\title{\texorpdfstring{$n$}{n}-Gon Correlators in \texorpdfstring{$d$}{d}-dimensions}

\author[a,b]{Yong-Xiang Su}
\author[a,b,c]{Dongyu Yang}
\affiliation[a]{University of Chinese Academy of Sciences, Beijing 100049, China}
\affiliation[b]{New Cornerstone Laboratory, Institute of Theoretical Physics, Chinese Academy of Sciences, Beijing 100190, China}
\affiliation[c]{School of Fundamental Physics and Mathematical Sciences, Hangzhou Institute for Advanced Study, UCAS, Hangzhou 310024, China}
\emailAdd{suyongxiang26@mails.ucas.ac.cn}
\emailAdd{yangdongyu24@mails.ucas.ac.cn}

\abstract{We study the analytic structure of scalar one-loop cosmological $n$-gon correlators using an integrand-level reduction and a Baikov representation adapted to linear tube-energy denominators. At spatial dimension $d=n$, we determine the algebraic factors that normalize their local leading singularities to unity and construct explicit $d\log$ forms for arbitrary $n$. For the complete box at $d=n=4$, we evaluate an external-energy discontinuity in a collinear kinematic family and find a nonvanishing complete elliptic integral of the second kind after summing all time orderings. We also compare successive tube-energy discontinuities in $d=3$ and $d=4$, and relate selected tube cuts to the cosmological dressing
representation. These results illustrate the importance of integration domains in determining the analytic structure beyond local leading singularities.}

\begin{document}
\maketitle

\section{Introduction}\label{sec:intro}

The analytic structure of cosmological correlators is
strongly constrained by locality, unitarity and symmetry~\cite{Arkani-Hamed:2017fdk,Arkani-Hamed:2018kmz,
Baumann:2019oyu,Anninos:2014lwa}.
At loop level, these constraints are less readily translated
into explicit results: both the interaction times and the
internal spatial momenta must be integrated, and the resulting
functions can have a considerably richer analytic structure
than their integrands~\cite{Benincasa:2022gtd,Lee:2023loops}.
Understanding this distinction is one of the motivations
for studying loop correlators directly.

In this work, we consider a massless scalar field in $(d+1)$-dimensional Lorentzian spacetime, with mostly-plus metric signature and general non-derivative interactions
\begin{equation}\label{eq:intro_lagrangian}
 \mathcal L=
 -\frac12\partial_\mu\phi\,\partial^\mu\phi
 -\sum_{m\geq3}\frac{\lambda_m}{m!}\phi^m.
\end{equation}
After Wick rotation, the Euclidean mixed-representation bulk-to-bulk propagator and the normalized propagator connecting an interaction vertex to the observation slice $t=0$ are
\begin{equation}\label{eq:intro_propagators}
 G_q(t,t')=\frac{e^{-q|t-t'|}}{2q},
 \qquad
 K_p(t)=\frac{G_p(t,0)}{G_p(0,0)}
 =e^{-p|t|}.
\end{equation}
We refer to the latter as the bulk-to-boundary factor.
For a one-loop $n$-gon, the $i$th vertex carries
total external energy $X_i=\sum_{a\in\mathcal E_i}|\vec p_a|$
and spatial momentum
$\vec k_i=\sum_{a\in\mathcal E_i}\vec p_a$, where $\vec p_a$ denotes the momentum of each external leg $a$ and $\mathcal{E}_i$ denotes the set of external legs connected with the $n$-gon vertex.
The number $n$ counts vertices around the loop,
each of which may have arbitrary valence.
With couplings, symmetry factors and external-leg
normalizations stripped off, the corresponding
equal-time correlator is the full-time integral
\eqref{eq:kernel}. We treat the $X_i$ as independent
analytic variables while imposing spatial momentum
conservation. The same free kernels arise after
conformal rescaling in cosmological settings,
although general interactions may introduce
additional time-dependent vertex factors.

This full-time representation is particularly useful
for in-in correlators, whose analytic properties
can be simpler than those of cosmological wavefunction
coefficients~\cite{Chowdhury:2023arc,
Arkani-Hamed:2025mce,Chowdhury:2026dwm}.
The connection to flat-space Feynman integrals
has been developed through cosmological dressing
rules~\cite{Chowdhury:2025ohm}, while tropical
and stringy integrals offer a complementary
description of their combinatorics~\cite{He:2026rpw,Zhu:2026xil}.

Various approaches have been developed to organize
cosmological integrals before spatial loop integration.
Kinematic flow relates differential equations and
singularities to graph tubings~\cite{Arkani-Hamed:2023kig,Arkani-Hamed:2023bsv,
He:2024olr,Baumann:2024mvm,Hang:2024xas},
while integration-by-parts and cohomological methods
provide complementary tools~\cite{Chen:2023iix,De:2023xue}.
For ordinary Feynman integrals, integrand reduction
and Baikov representations~\cite{Ossola:2006us,Baikov:1996iu,Frellesvig:2017aai},
together with leading singularities and $d\log$ forms~\cite{Cachazo:2008vp,Henn:2013pwa,Herrmann:2019upk},
have proved especially effective.
Their relation to hyperbolic geometry~\cite{Davydychev:1997wa,Schnetz:2010pd,
Bourjaily:2019dix} further suggests a geometric
description of integrated answers.
Recent evaluations of cosmological bubble and
triangle correlators~\cite{Pimentel:2026kqc,
Henn:2026lfz}, and elliptic structures in
one-loop wavefunction integrals~\cite{Benincasa:2024ptf}, provide useful
points of comparison.

In this paper, we first reduce the complete scalar
$n$-gon to $n\,n!/2$ basic-integral terms,
organized into $n2^{n-2}$ groups with common
rational coefficients. Each basic integral contains
$n-1$ linear tube-energy denominators.
This is an algebraic reduction before loop integration,
rather than a reduction to independent master integrals.
We then develop a Baikov representation adapted
to these tube denominators. Their incidence structure
makes the Baikov polynomial quadratic in one remaining
loop energy. At $d=n$, its integration yields a
square-root measure, for which we determine the
unit local leading-singularity normalization
and construct explicit $d\log$ forms for arbitrary $n$.

The basic bubble and triangle at $d=n=2,3$ admit
geometric interpretations in terms of an analytically
continued hyperbolic area and a signed sum of
ideal-tetrahedron volumes, respectively.
For the box at $d=n=4$, we organize the 48 basic
terms into three inequivalent routing patterns
and investigate the analytic structure of the
complete correlator through its energy discontinuities.
We first study a large-energy limit, where the
threshold discontinuity reduces to a one-dimensional
integral associated with a genus-two curve at
generic spatial kinematics, with elliptic periods
appearing in a further limit.
More explicitly, in a collinear configuration we
evaluate a discontinuity of the full correlator
in terms of the complete elliptic integral of
the second kind, whose contribution survives
the sum over all time orderings.
We then use the Baikov representation to compare
successive tube-energy discontinuities of the
basic box in $d=3$ and $d=4$, showing how
non-polylogarithmic dependence can arise despite
algebraic maximal residues and a $d\log$ form.
Finally, we relate selected tube-energy cuts to
generalized cuts in the cosmological dressing
representation, reproducing both the reduction
coefficient and the local leading singularity.
These results connect the analytic structure of
the full correlator with cosmological cutting rules
and dispersive methods~\cite{Goodhew:2020hob,
Melville:2021lst,Goodhew:2021oqg,
Chowdhury:2026upp,Colipi-Marchant:2025oin,
Das:2025qsh}.

Section~\ref{sec:reduction} presents the integrand
reduction and takes the box as a concrete example.
Section~\ref{sec:baikov} develops the Baikov
representation, followed by the leading singularities in Section~\ref{sec:canonical}
and $d\log$ forms in Section~\ref{sec:dform}.
Section~\ref{sec:box} analyzes the box under large energy scales and its discontinuities, both for the whole result and for the single basic integral.
The appendices contain the $d=n-1$ Baikov limit,
the bubble and triangle evaluations and their
geometric interpretations, and additional
kinematic limits.

\section{Integrand reduction for the \texorpdfstring{$n$}{n}-gon correlator}\label{sec:reduction}

We begin with the bubble and triangle, which illustrate how the time-ordered contributions reduce to a smaller set of spatial loop integrals. The same partial fraction identity then gives the general $n$-gon expansion.

\begin{figure}[!htbp]
 \centering
 \includegraphics[width=\textwidth]{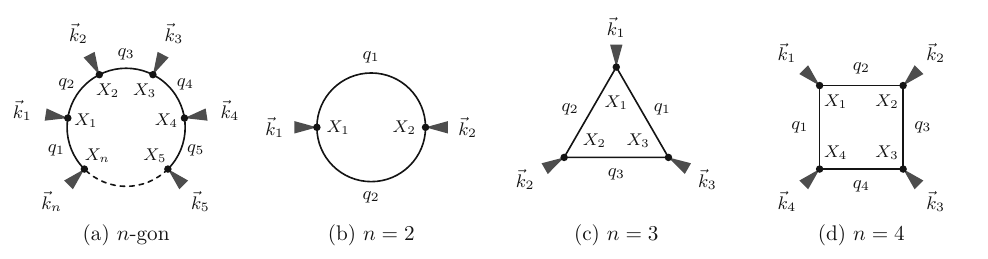}
 \caption{One-loop polygons with interaction vertices of arbitrary valence.
Each shaded wedge represents the external legs attached to vertex $i$,
with total spatial momentum $\vec k_i=\sum_{a\in\mathcal E_i}\vec p_a$
and total energy $X_i=\sum_{a\in\mathcal E_i}\omega_a$.
A vertex with $m_i=|\mathcal E_i|$ external legs has valence $m_i+2$;
$n$ counts loop vertices. The internal energy $q_i$ labels edge $(i-1,i)$,
with cyclic indices. The dashed arc in (a) represents omitted vertices
and propagators.}
 \label{fig:ngon_graphs}
\end{figure}

Consider the scalar one-loop kernel
\begin{equation}\label{eq:kernel}
 \mathcal I_n=
 \int\dd^d\vec \ell\int_{\mathbb R^n}\dd t_1\cdots\dd t_n\,
 \prod_{i=1}^n e^{-X_i|t_i|}
 \prod_{i=1}^n\frac{e^{-q_i|t_i-t_{i-1}|}}{2q_i},
 \qquad t_0=t_n,
\end{equation}
with the momentum routing
\begin{equation}\label{eq:momentum_routing}
 q_i=\left|\vec \ell+\sum_{a=1}^{i-1}\vec k_a\right|,
 \qquad q_1=|\vec \ell|,
 \qquad \sum_{i=1}^n\vec k_i=0.
\end{equation}
All external momenta are incoming. Edge $i$ joins vertices $i-1$ and $i$ cyclically; the bubble has two distinct edges connecting its vertices, as shown in Figure~\ref{fig:ngon_graphs}. We initially take $X_i>0$ and real Euclidean spatial momenta, treating the $X_i$ as independent of the conserved spatial momenta. Couplings, external-leg factors, symmetry factors and the momentum-conservation delta function are omitted, and the loop measure contains no factor of $(2\pi)^{-d}$. The integrated identities below are understood as being evaluated in a convergent domain or with a common regulator (we provide some examples of divergent integrals in Appendix~\ref{app:limits}).

\begin{figure}[htbp]
 \centering
 \includegraphics[width=.98\textwidth]{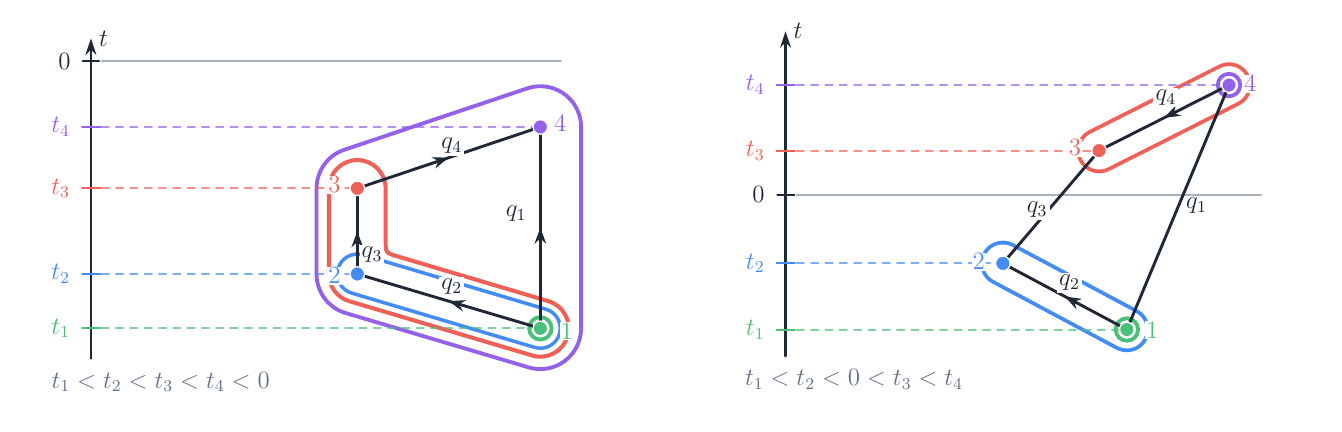}
 \caption{Two representative box time orderings, $t_1<t_2<t_3<t_4<0$ (left) and $t_1<t_2<0<t_3<t_4$ (right). Including their time reverses gives $I_{1,2,3,4}$ and $I_{1,2;4,3}$. Successive tubes enclose initial segments of the time-ordered vertices, and the edges crossing them determine their energy denominators.}
 \label{fig:time_sectors}
\end{figure}

Each ordering of the interaction times relative to one another and to $t=0$ determines a tubing structure, as shown in Figure \ref{fig:time_sectors}. A tube enclosing vertices $i_1,\ldots,i_r$ contributes the energy denominator
\begin{equation}\label{eq:tubes}
 L_{i_1\cdots i_r}=X_{i_1}+\cdots+X_{i_r}
 +\sum_{e\in\partial\{i_1,\ldots,i_r\}}q_e,
\end{equation}
where $\partial\{i_1,\ldots,i_r\}$ consists of edges with exactly one endpoint among the enclosed vertices. The enclosed vertices need not be connected. In particular, $L_{1\cdots n}=\sum_{i=1}^nX_i$.

For each side of $t=0$, list the vertices from the farthest to the closest time. Integrating the successive time intervals gives a denominator for every initial segment of each list. Combining a time ordering with its time reverse, we write
\begin{align}\label{eq:tubing_integral}
I_{a_1,\ldots,a_r;b_1,\ldots,b_s}^{(n)}
 =\int\frac{\dd^d\vec \ell}{2^{n-1}q_1\cdots q_n}
 &\frac{1}{L_{a_1}L_{a_1a_2}\cdots L_{a_1\cdots a_r}}
 \frac{1}{L_{b_1}L_{b_1b_2}\cdots L_{b_1\cdots b_s}}.
\end{align}
An empty list contributes no denominators and is omitted from the subscript. The factor $2$ from time reversal is included in the overall $2^{1-n}$, so $I_{a_1,\ldots,a_r;b_1,\ldots,b_s}^{(n)}=I_{b_1,\ldots,b_s;a_1,\ldots,a_r}^{(n)}$. Initially the two lists contain all $n$ vertices; after reduction they contain $n-1$. We suppress the superscript $(n)$ in the following low-point examples.

For the bubble, $L_1=X_1+q_1+q_2$ and $L_2=X_2+q_1+q_2$. Its three paired time orderings reduce immediately to
\begin{equation}
\begin{aligned}
\label{eq:bubble_reduction}
 \mathcal I_2=&I_{1,2}+I_{2,1}+I_{1;2}
 =\frac{I_1+I_2}{X_1+X_2}+\frac{I_1-I_2}{X_2-X_1}\\=&\left(\frac1{X_1+X_2}-\frac1{X_1-X_2}\right)I_1
  +\left(\frac1{X_1+X_2}-\frac1{X_2-X_1}\right)I_2.
\end{aligned}
\end{equation}
Thus only two basic integrals, each with one tube denominator, are needed.

Likewise, the triangle has twelve paired time orderings, which reduce to nine basic-integral terms. Grouping terms with equal coefficients gives the compact expression
\begin{equation}
\label{eq:triangle_reduction}
 \mathcal I_3=\sum_{\mathrm{cyc}(i,j,k)}\Bigg[
 \left(\frac1{X_1+X_2+X_3}-\frac1{X_i+X_j-X_k}\right)
 (I_{i,j}+I_{j,i})+\frac{2X_k}{X_k^2-(X_j-X_i)^2}\,I_{i;j}\Bigg],
\end{equation}
where the sum runs over $(i,j,k)=(1,2,3),(2,3,1),(3,1,2)$. The nine integrals fall into six groups. Their integrated forms will be discussed below, including the comparison with~\cite{Henn:2026lfz}.

For the box, the representative time-ordering regions $t_1<t_2<t_3<t_4<0$ and $t_1<t_2<0<t_3<t_4$, together with their time reverses, give 
\begin{align}\label{eq:time_box_example}
 I_{1,2,3,4}
 &=\int\frac{\dd^d\vec \ell}{8q_1q_2q_3q_4}
 \frac{1}{L_1L_{12}L_{123}L_{1234}},\qquad
 I_{1,2;4,3}
 =\int\frac{\dd^d\vec \ell}{8q_1q_2q_3q_4}
 \frac{1}{L_1L_{12}L_4L_{34}}.
\end{align}
Here $L_1=X_1+q_1+q_2$, $L_{12}=X_1+X_2+q_1+q_3$, $L_{123}=X_1+X_2+X_3+q_1+q_4$ and $L_{1234}=X_1+X_2+X_3+X_4$. The time orderings are illustrated in Figure~\ref{fig:time_sectors}. Both are reduced by the same rule, and the full box expansion is given in Section~\ref{sec:box_expansion}.

The reduction in these examples relies only on the fact that complementary sets of vertices have the same boundary edges. For two nonempty complementary lists $a_1,\ldots,a_m$ and $b_1,\ldots,b_{n-m}$, we have
\begin{align}\label{eq:partial_fraction}
 \frac1{L_{a_1\cdots a_m}L_{b_1\cdots b_{n-m}}}
 =\frac{1}{\displaystyle\sum_{r=1}^{n-m}X_{b_r}-\sum_{r=1}^mX_{a_r}}
 \left(\frac1{L_{a_1\cdots a_m}}-\frac1{L_{b_1\cdots b_{n-m}}}\right).
\end{align}
Consequently, an integral with all $n$ vertices \eqref{eq:tubing_integral} distributed between the two lists satisfies
\begin{align}\label{eq:capping}
 I_{a_1,\ldots,a_m;b_1,\ldots,b_{n-m}}^{(n)}
 =\frac{
 I_{a_1,\ldots,a_m;b_1,\ldots,b_{n-m-1}}^{(n)}
 -I_{a_1,\ldots,a_{m-1};b_1,\ldots,b_{n-m}}^{(n)}
 }{\displaystyle\sum_{r=1}^{n-m}X_{b_r}-\sum_{r=1}^{m}X_{a_r}}.
\end{align}
If all vertices lie on one side of zero, the last tube instead encloses the whole graph and contributes simply $1/\sum_iX_i$. In either case, one denominator can be removed. We call the resulting integrals with $n-1$ tube denominators basic integrals. This reduction takes place at the level of the rational integrand, before spatial loop integration.

For arbitrary $n$, the original time-ordering expansion can be expressed using only permutations of the vertices
\begin{equation}\label{eq:time_integral}
 \mathcal I_n=\frac12\sum_{\pi\in S_n}\sum_{m=0}^n
 I_{\pi(1),\ldots,\pi(m);\pi(m+1),\ldots,\pi(n)}^{(n)}.
\end{equation}
The two lists describe vertices on opposite sides of zero, each ordered from the farthest to the closest time; either list can be empty. Interchanging the lists amounts to time reversal and explains the factor $1/2$. Thus there are $(n+1)!/2$ paired time-ordering terms.

After using \eqref{eq:capping}, each basic integral arises from two ways of adjoining its omitted vertex to one of the two lists. Their rational coefficients combine into
\begin{align}\label{eq:reduction}
 \mathcal I_n={}&\frac12\sum_{\pi\in S_n}\sum_{m=0}^{n-1}
 \left(
 \frac{1}{\displaystyle\sum_{i=1}^nX_i-2\sum_{r=1}^{m}X_{\pi(r)}}
 -\frac{1}{\displaystyle\sum_{i=1}^nX_i-2\sum_{r=1}^{m+1}X_{\pi(r)}}
 \right)
I_{\pi(1),\ldots,\pi(m);\pi(m+2),\ldots,\pi(n)}^{(n)}.
\end{align}
When we encounter empty sums, they vanish, and empty lists contribute no denominators. The factor $1/2$ again removes the duplication under exchange of the two lists. In particular, the coefficient of a basic integral depends only on the omitted vertex $\pi(m+1)$ and the set of vertices $\{\pi(1),\ldots,\pi(m)\}$, not on their ordering. Exchanging that set with the complementary set of the remaining vertices gives the same coefficient. This is the grouping of basic integrals into sectors, the formula reproduces \eqref{eq:bubble_reduction} and \eqref{eq:triangle_reduction}.

We note that this decomposition looks akin to the causal decomposition in the existing contexts \cite{Capatti:2020ytd,Kromin:2022txz}, which tries to integrate out the time component of the loop momentum $\ell^0$ for the $n$-gon amplitude. A related pole structure also appears in the cosmological
dressing representation~\cite{Chowdhury:2025ohm},
as briefly illustrated for the box in Section~\ref{sec:box}.

Here we show an explicit example for boxes.
\paragraph{Example: Box}
\label{sec:box_expansion}
Applying the partial-fraction identity
\eqref{eq:capping} reduces these contributions
to basic integrals with three tube
denominators. 
The resulting expression therefore contains
\begin{equation}
N_{\rm basic}
=4\times3!+4\times3\times2!
=48
\end{equation}
basic-integral terms, grouped into
$4+12=16$ combinations with common
rational coefficients, as listed in Table \ref{tab:box_basic_groups}.

\begin{table}[htbp]
\centering
\footnotesize
\begin{tblr}{
colspec={c c X[l]},
row{1}={font=\bfseries},
row{5}={abovesep=4pt}
}
\toprule
Omitted vertex
& Singleton vertex
& Basic-integral combination\\
\midrule
$1$ & -- &
$I_{2,3,4}+I_{2,4,3}
+I_{3,2,4}+I_{3,4,2}
+I_{4,2,3}+I_{4,3,2}$\\

$2$ & -- &
$I_{1,3,4}+I_{1,4,3}
+I_{3,1,4}+I_{3,4,1}
+I_{4,1,3}+I_{4,3,1}$\\

$3$ & -- &
$I_{1,2,4}+I_{1,4,2}
+I_{2,1,4}+I_{2,4,1}
+I_{4,1,2}+I_{4,2,1}$\\

$4$ & -- &
$I_{1,2,3}+I_{1,3,2}
+I_{2,1,3}+I_{2,3,1}
+I_{3,1,2}+I_{3,2,1}$\\
\midrule

$2$ & $1$ &
$I_{1;3,4}+I_{1;4,3}$\\
$3$ & $1$ &
$I_{1;2,4}+I_{1;4,2}$\\
$4$ & $1$ &
$I_{1;2,3}+I_{1;3,2}$\\

$1$ & $2$ &
$I_{2;3,4}+I_{2;4,3}$\\
$3$ & $2$ &
$I_{2;1,4}+I_{2;4,1}$\\
$4$ & $2$ &
$I_{2;1,3}+I_{2;3,1}$\\

$1$ & $3$ &
$I_{3;2,4}+I_{3;4,2}$\\
$2$ & $3$ &
$I_{3;1,4}+I_{3;4,1}$\\
$4$ & $3$ &
$I_{3;1,2}+I_{3;2,1}$\\

$1$ & $4$ &
$I_{4;2,3}+I_{4;3,2}$\\
$2$ & $4$ &
$I_{4;1,3}+I_{4;3,1}$\\
$3$ & $4$ &
$I_{4;1,2}+I_{4;2,1}$\\
\bottomrule
\end{tblr}
\caption{The 16 combinations of basic box integrals.
The first four contain six orderings each,
while the remaining twelve contain two.
The entries count terms in the reduction,
not independent integrated functions.}
\label{tab:box_basic_groups}
\end{table}

Although the reduction contains 48 basic
terms, their tube-energy denominators
have only three inequivalent routing
patterns under the dihedral symmetry
of the box. The dihedral transformations
act simultaneously on the vertex labels,
external energies and momenta, and can
be implemented by shifts or reflections
of the loop momentum. Consequently,
they preserve the loop measure and relate
integrals with relabeled external data.

\begin{figure}[htbp]
\centering
\IfFileExists{Graphs/box_tubings.pdf}{
  \includegraphics[scale=1]{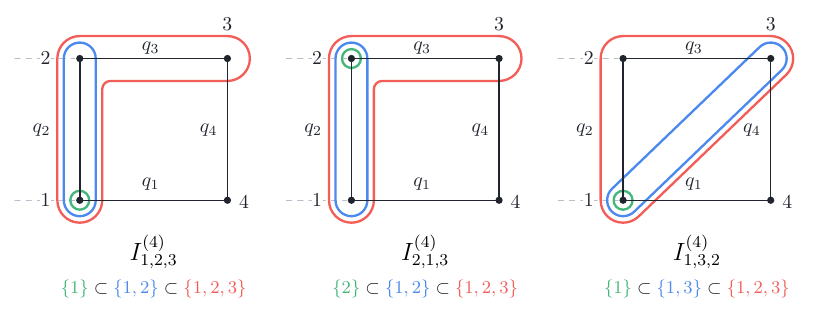}
}{
  \fbox{\parbox{0.82\linewidth}{\centering
  The three box tubing diagrams are included
  from \texttt{Graphs/box\_tubings.pdf} in the full manuscript.}}
}
\caption{
Three inequivalent box tubing structures under
dihedral relabeling. The red contours enclose
$\{1,2,3\}$ in all three diagrams.
The blue contours enclose $\{1,2\}$ in the first
two diagrams and $\{1,3\}$ in the third.
The green contours enclose $\{1\}$, $\{2\}$,
and $\{1\}$, respectively.
The three structures correspond to the basic
integrals $I_{1,2,3}$, $I_{2,1,3}$,
and $I_{1,3,2}$.
}
\label{fig:box_routing_templates}
\end{figure}

We may choose the following three
one-sided basic integrals as representatives
\begin{align}
&I_{1,2,3}
=\int\frac{\dd^d\vec\ell}{8q_1q_2q_3q_4}
\frac{1}{L_1L_{12}L_{123}},
\quad
I_{2,1,3}
=\int\frac{\dd^d\vec\ell}{8q_1q_2q_3q_4}
\frac{1}{L_2L_{12}L_{123}},
\notag\\&
I_{1,3,2}
=\int\frac{\dd^d\vec\ell}{8q_1q_2q_3q_4}
\frac{1}{L_1L_{13}L_{123}}.
\label{eq:box_routing_representatives}
\end{align}
The corresponding tubing structures
are shown in Figure~\ref{fig:box_routing_templates}.
In the first integral, the three tubes
enclose the successive subsets
$\{1\}$, $\{1,2\}$ and $\{1,2,3\}$.
The second has the same two outer tubes
but begins with $\{2\}$. In the third,
the intermediate tube encloses
the disconnected subset $\{1,3\}$,
whose boundary crosses all four
internal edges.

Each routing pattern also occurs for
split basic integrals. For example,
$I_{1;4,3}$, $I_{2;4,3}$ and
$I_{1;4,2}$ have the same respective
sets of boundary-energy contributions
as the three integrals in
\eqref{eq:box_routing_representatives},
up to the ordering of their denominators.
The vertex-energy sums, however, differ.
Thus a routing pattern alone does not
completely specify a basic integral.

Under the $D_4$ action, the 48 labelled
basic terms form six orbits of eight
terms each, with two orbits associated
with each of the three routing patterns.
The detailed relabelings are collected
in Appendix~\ref{app:routing}.
This classification concerns the
denominator structures and does not
imply linear independence of the
corresponding integrated functions.

There is, in addition, a useful partial-fraction
relation between the second and third routing
representatives. Here $A_1,A_2,A_3$ denote the
vertex-energy parts of the three tube denominators in the
chosen routing. At the integrand level the relation follows
from
\begin{align}
&\frac{1}{(A_1+q_2+q_3)(A_2+q_1+q_3)(A_3+q_1+q_4)}
=
\frac{1}{(A_1+q_2+q_3)(A_2+q_1+q_3)
(A_1+A_3+q_1+q_2+q_3+q_4)}
\notag\\
&\qquad+
\frac{1}{(A_3+q_1+q_4)(A_2+q_1+q_3)
(A_1+A_3+q_1+q_2+q_3+q_4)}.
\label{eq:box_bonus_pf}
\end{align}
After the corresponding shifts or reflections of the loop
momentum, this gives
\begin{align}
I_{2,1,3}
(A_1,A_2,A_3;\vec k_1,\vec k_2,\vec k_3,\vec k_4)
&=
I_{1,3,2}
\!\left(
A_1,A_1+A_3,A_2;
-\vec k_2,\vec k_2+\vec k_3,\vec k_4,\vec k_1+\vec k_2
\right)
\notag\\
&+
I_{1,3,2}
\!\left(
A_3,A_1+A_3,A_2;
-\vec k_4,-\vec k_2-\vec k_3,\vec k_2,-\vec k_1-\vec k_2
\right).
\label{eq:box_bonus_relation}
\end{align}
Thus the second routing representative need not be kept as a
separate seed: it is expressible through two instances of the
third routing type with shifted energy arguments and relabeled
external momenta. The three dihedral routing classes therefore
reduce to two types of basic integral, without changing the
48-term expansion or the 16 groups in Table~\ref{tab:box_basic_groups}.

\section{Baikov representation}\label{sec:baikov}
We now express the basic integrals of \eqref{eq:reduction} in terms
of internal energies. Their Baikov representation allows one energy
integration to be performed explicitly, particularly simple at $d=n$.

Consider any basic integral from Section~\ref{sec:reduction}. Its $n-1$
tube denominators are linear in $q_1,\ldots,q_n$, so we may write
\begin{equation}\label{eq:z_variables}
 I^{(n)}=\int\frac{\dd^d\vec \ell}{2^{n-1}q_1\cdots q_n}
 \prod_{i=1}^{n-1}\frac1{z_i},\qquad
 z_i=A_i+\sum_{j=1}^nc_{ij}q_j,
\end{equation}
where $A_i$ is the enclosed vertex-energy sum and $c_{ij} = 1$ when edge $j$ crosses that tube and zero otherwise. Thus $z_i$ are the tube energies $L_{i_1\cdots i_r}$ of \eqref{eq:tubes}, in the
order of the chosen basic integral. To see it more clearly, consider the box example, $I_{1,2,3}^{(4)}$, with the tubing structure shown in left-most figure of \ref{fig:box_routing_templates}, whose $z$'s and $c$'s are
\begin{align}
    z_1=L_1,\qquad z_2=L_{1,2},\qquad z_3=L_{1,2,3},\qquad c=\begin{pmatrix}
        1&1&0&0\\1&0&1&0\\1&0&0&1
    \end{pmatrix}.
\end{align}
We now make the following change of variables by replacing the individual cartesian components of $\vec q$, i.e., $q^{(1)}, \cdots, q^{(d)}$ by the tube energies $z_1, \cdots, z_{n-1}$. This also explains the restriction $d \geq n$\footnote{The case $d = n -1$ also admits a Baikov representation but it requires a separate treatment which will be addressed in Appendix \ref{app:details_Baikov}.}. With this change of variables,  the Jacobian corresponding to the transformation becomes

\begin{equation}\label{eq:change_of_variable}
 \dd^d\vec\ell=
 \frac{
 2\pi^{(d-n+1)/2}\mathcal B^{(d-n-1)/2}
 q_1\cdots q_n
 }{
 \Gamma\!\left(\frac{d-n+1}{2}\right)
 (\det H)^{(d-n)/2}
 \left|\det(c_{i,j+1})_{i,j=1}^{n-1}\right|
 }
 \dd q_1\,\dd z_1\cdots\dd z_{n-1}.
\end{equation}
This leaves us with the following representation for \eqref{eq:z_variables}. By a direct calculation it can be shown that $\mathcal B$ is quadratic in $q_1$ (where the polynomial is given in detail in Appendix \ref{app:details_Baikov}) and generally $\left|\det(c_{i,j+1})_{i,j=1}^{n-1}\right|=1$ which allows us to integrate it out. This leaves us with 
\begin{equation}\label{eq:integrated_baikov}
 I^{(n)}=
 \frac{2^{2-n}\pi^{(d-n+1)/2}
 \Gamma\bigl(\frac{d-n+1}{2}\bigr)}
 {\Gamma(d-n+1)(\det H)^{(d-n)/2}}
 \int_\Sigma\frac{\dd z_1\cdots\dd z_{n-1}}
 {z_1\cdots z_{n-1}}
 \frac{\Delta(z)^{(d-n)/2}}
 {\Omega(z)^{(d-n+1)/2}}.
\end{equation}
The definition of polynomials and the integration are discussed in details in Appendix \ref{app:details_Baikov}.

At $d=n$, the discriminant drops out of the integrand, and the
representation takes the particularly simple form
\begin{equation}\label{eq:reduced}
\displaystyle
 I^{(n)}\big|_{d=n}=2^{2-n}\pi\,\int_\Sigma
 \frac{\dd z_1\cdots\dd z_{n-1}}
 {\sqrt{\Omega(z)}\,z_1\cdots z_{n-1}}.
\end{equation}
The square root is positive on the real domain. Although
$\Delta(z)$ drops out of the integrand, $\Delta(z)\geq0$ still
constrains $\Sigma$; this matters for the analytic structure.

%%%%%%%%%%%%%%%%%%%%%%%%%%%%
%%%%%%%%%%%%%%%%%%%%%%%%%%%%
\subsection{Unit leading singularities \texorpdfstring{$d=n$}{d=n}}\label{sec:canonical}

The Baikov representation also makes it natural to study discontinuities in the external-energy sums $A_i$ of \eqref{eq:z_variables}. Keeping the spatial momenta and the remaining $A_j$ fixed, the contribution from a tube denominator is
\begin{equation}\label{eq:tube_energy_disc}
 -\frac{1}{2\pi i}\Disc_{A_i}\frac{1}{z_i}
 =\delta(z_i),\qquad
 z_i=A_i+\sum_{j=1}^{n}c_{ij}q_j.
\end{equation}
Thus the discontinuity puts the corresponding tube energy on shell. This is analogous to the energy discontinuities appearing in cosmological cutting rules and dispersive constructions~\cite{Goodhew:2020hob,Goodhew:2021oqg,Chowdhury:2026upp}.

At $d=n$, the simultaneous contribution from all $n-1$ tube poles follows from \eqref{eq:reduced}
\begin{align}\label{eq:sequential_tube_disc}
 &\left.\left[\prod_{i=1}^{n-1}\left(-\frac{\Disc_{A_i}}{2\pi i}\right)I^{(n)}\right]\right|_{\substack{d=n\\\text{tube poles}}}
 =\int\frac{\dd^n\vec \ell}{2^{n-1}q_1\cdots q_n}
 \prod_{i=1}^{n-1}\delta(z_i)
 =\frac{2^{2-n}\pi}{\sqrt{\Omega(0)}}.
\end{align}
The delta-function expression applies when the constraints have support on the real integration domain; otherwise the last expression is understood as the analytic continuation of the local tube-pole contribution.

For a chosen basic integral, the algebraic factor in \eqref{eq:sequential_tube_disc} is determined by
\begin{equation}\label{eq:LS_root}
 \Omega(0)
 =\sum_{a,b=1}^{n-1}\sigma_a\sigma_b\,y_a(0)(\adj H)_{ab}y_b(0)-\det H
 =-\det\!\left[H-\bigl(\sigma_a y_a(0)\,\sigma_b y_b(0)\bigr)_{a,b=1}^{n-1}\right].
\end{equation}
The two signs of the square root correspond to the two local sheets. In particular, the normalized differential form has unit residues
\begin{equation}\label{eq:local_LS}
 \left.\operatorname*{Res}_{z_1=\cdots=z_{n-1}=0}
 \frac{\sqrt{\Omega(0)}\,\dd z_1\wedge\cdots\wedge\dd z_{n-1}}
 {\sqrt{\Omega(z)}\,z_1\cdots z_{n-1}}\right|_{\pm}=\pm1.
\end{equation}
Here the residue is taken in the order $z_1,\ldots,z_{n-1}$, fixing the orientation. We will construct the corresponding $d\log$ form in the next subsection.

One consequence of \eqref{eq:LS_root} is that the normalization is common to all orderings within each group in \eqref{eq:reduction}. At $z=0$, consecutive differences of the tube equations fix the components $\sigma_a y_a(0)$, up to an overall sign, by the omitted vertex and the unordered division of the remaining vertices. Permuting the vertices within either list therefore leaves $\Omega(0)$ unchanged. Thus the $n2^{n-2}$ groups have common square-root factors, generically distinct between groups, despite the $n\,n!/2$ basic-integral terms.
%%%%%%%%%%%%%%%%%%%%%%%%%%%%
%%%%%%%%%%%%%%%%%%%%%%%%%%%%
\subsection{An explicit \texorpdfstring{$d\log$}{log} form for \texorpdfstring{$d = n$}{d=n}}\label{sec:dform}
The quadratic nature of $\Omega(z)$ also allows the square root in
\eqref{eq:reduced} to be absorbed into logarithmic differentials.
For a fixed basic integral, choose a branch of $\sqrt{\Omega(0)}$
with $\Omega(0)\neq0$ and define
\begin{equation}\label{eq:F_common}
 F(z)=\sqrt{\Omega(0)}\sqrt{\Omega(z)}+\Omega(0)
 +\frac12\sum_{i=1}^{n-1}z_i
 \left.\frac{\partial\Omega}{\partial z_i}\right|_{z=0}.
\end{equation}
%Here the concrete proof is commented.
\begin{comment}
   This is equivalent to the expression obtained from
\eqref{eq:quadratic_baikov}. All differentials are taken at fixed external energies and momenta. The identity follows directly from the quadratic dependence of
$\Omega$ on the $z_i$. In particular,
\begin{equation}
 \sum_{i=1}^{n-1}z_i\frac{\partial\Omega(z)}{\partial z_i}
 =2\bigl[\Omega(z)-\Omega(0)\bigr]
 -\sum_{i=1}^{n-1}z_i
 \left.\frac{\partial\Omega}{\partial z_i}\right|_{z=0}.
\end{equation}
Differentiating \eqref{eq:F_common} and using this relation gives
\begin{align}
 \sum_i z_i\frac{\partial F}{\partial z_i}
 & =\frac{\sqrt{\Omega(0)}}{2\sqrt{\Omega(z)}}
 \left[2\bigl(\Omega(z)-\Omega(0)\bigr)
 -\sum_i z_i\left.\frac{\partial\Omega}{\partial z_i}\right|_{z=0}\right]
 +\frac12\sum_i z_i\left.\frac{\partial\Omega}{\partial z_i}\right|_{z=0}
=\left(1-\frac{\sqrt{\Omega(0)}}{\sqrt{\Omega(z)}}\right)F(z).
\end{align}
The matrix determinant lemma applied to the logarithmic differentials
then yields
\begin{align}
 \bigwedge_{i=1}^{n-1}\dd\log\frac{z_i}{F(z)}
 &=\left(1-\frac1{F(z)}\sum_i z_i\frac{\partial F}{\partial z_i}\right)
 \frac{\dd z_1\wedge\cdots\wedge\dd z_{n-1}}{z_1\cdots z_{n-1}}
=\frac{\sqrt{\Omega(0)}}{\sqrt{\Omega(z)}}
 \frac{\dd z_1\wedge\cdots\wedge\dd z_{n-1}}{z_1\cdots z_{n-1}}.
\end{align} 
\end{comment}
Then each basic integral at $d=n$ becomes
\begin{equation}\label{eq:normalized_basic}
 I^{(n)}\big|_{d=n}
 =\frac{2^{2-n}\pi}{\sqrt{\Omega(0)}}
 \int_\Sigma\bigwedge_{i=1}^{n-1}\dd\log\frac{z_i}{F(z)}.
\end{equation}
\begin{comment}
    Substituting this into the reduction \eqref{eq:reduction} gives
\begin{align}\label{eq:normalized_assembly}
 \mathcal I_n\big|_{d=n}
 ={}&\frac{2^{2-n}\pi}{2}\sum_{\pi\in S_n}\sum_{m=0}^{n-1}
 \left(
 \frac{1}{\displaystyle\sum_{i=1}^n X_i-2\sum_{r=1}^m X_{\pi(r)}}
 -\frac{1}{\displaystyle\sum_{i=1}^n X_i-2\sum_{r=1}^{m+1}X_{\pi(r)}}
 \right)\frac1{\sqrt{\Omega(0)}}
 \int_\Sigma\bigwedge_{i=1}^{n-1}\dd\log\frac{z_i}{F(z)}.
\end{align}
\end{comment}
Here $\Omega$, $F$, $z_i$ and $\Sigma$ refer to the basic integral
with ordered lists $(\pi(1),\ldots,\pi(m))$ and
$(\pi(m+2),\ldots,\pi(n))$. The same branch of $\sqrt{\Omega(0)}$
may be chosen for all orderings in a group. These identities concern
the differential forms, rather than their integration domains;
in particular, a $d\log$ form by itself does not guarantee an
evaluation in multiple polylogarithms or a hyperbolic-volume
interpretation~\cite{Duhr:2020gdd}.

\section{The box at \texorpdfstring{$d=n=4$}{d=n=4}}\label{sec:box}
We now turn to the box in four spatial dimensions. The local leading singularities of the basic integrals admit algebraic normalizations to unity, and the corresponding normalized differential forms have the $d\log$
representation of Section~\ref{sec:dform}. This does not, however, fix the class of functions obtained after integration. We will study the complete box, including all its time orderings, and show that an elliptic integral
survives in one of its energy discontinuities. We then return to the
Baikov representation to see why this is compatible with the simple
form of the basic integrands. Throughout this section we write
$\mathcal I_4=\mathcal I_4^{(4)}$ and omit the superscript $(4)$ on the
basic box integrals.

\paragraph{Large-energy limit of the complete box.}
We begin directly with the time-integral representation \eqref{eq:kernel}, keeping the spatial momenta fixed and set $X_2=X_3=X_4=M$. Since $M e^{-M|t|}/2$ approaches $\delta(t)$ as $M\to\infty$, the last three time integrals set $t_2=t_3=t_4=0$. The remaining integral over $t_1$ gives
\begin{equation}\label{eq:box_large_energy_limit}
 \lim_{M\to\infty}M^3\mathcal I_4(X_1,M,M,M)
 =\int\frac{\dd^4\vec q_1}{q_1q_2q_3q_4}\,
 \frac{1}{X_1+q_1+q_2}.
\end{equation}

A useful way to probe \eqref{eq:box_large_energy_limit} is to continue $X_1$ through the threshold $X_1=-|\vec k_1|$. We define $\Disc_{X_1}f(-x)=f(-x+i0)-f(-x-i0)$, with $x>0$. The origin of the cut is already visible before taking $M$ large. For fixed $t_2,t_3,t_4$, the two ends of the $t_1$ integral behave as
$e^{-(X_1+q_1+q_2)|t_1|}e^{\pm(q_2t_2+q_1t_4)}$.
Thus the contribution associated with the pole $X_1+q_1+q_2=0$ is
\begin{align}\label{eq:box_X1_cut_before_limit}
 -\frac{\Disc_{X_1}\mathcal I_4}{2\pi i}\bigg|_{X_1=-x}
 ={}&\int\frac{\dd^4\vec q_1}{16q_1q_2q_3q_4}\,
 \delta(x-q_1-q_2)
 \int_{\mathbb R^3}\dd t_2\,\dd t_3\,\dd t_4\,
 e^{-X_2|t_2|-X_3|t_3|-X_4|t_4|}\notag\\[-2pt]
 &\quad\times e^{-q_3|t_3-t_2|-q_4|t_4-t_3|}
 \left[e^{q_2t_2+q_1t_4}+e^{-q_2t_2-q_1t_4}\right].
\end{align}
Here the other external energies are taken sufficiently large that no
other energy threshold is crossed in the range of $x$ under
consideration. Equation~\eqref{eq:box_X1_cut_before_limit} follows by
isolating the two large-$|t_1|$ tails of the original kernel. In the
large-$M$ limit both exponentials in square brackets tend to one, so
\begin{equation}\label{eq:box_large_energy_cut}
 \lim_{M\to\infty}M^3
 \left[-\frac{\Disc_{X_1}\mathcal I_4(-x,M,M,M)}{2\pi i}\right]
 =\int\frac{\dd^4\vec q_1}{q_1q_2q_3q_4}\,
 \delta(x-q_1-q_2),\qquad x>|\vec k_1|.
\end{equation}
Such external-energy discontinuities are closely related
to cosmological cutting and dispersive constructions
\cite{Goodhew:2021oqg,Chowdhury:2026upp}.

To examine the threshold, put $\vec k_1=k_1\hat n$ with $k_1>0$. The surface $q_1+q_2=x$ is a prolate ellipsoid for $x>k_1$; as $x\to k_1$ it contracts onto the segment $\vec q_1=-q_1\hat n$, $0\leq q_1\leq k_1$. Resolving the loop
momentum along $\hat n$ gives
\begin{equation}
 \frac{\dd^4\vec q_1}{q_1q_2}\,\delta(x-q_1-q_2)
 =\frac{\sqrt{(x^2-k_1^2)[k_1^2-(2q_1-x)^2]}}
 {2k_1^2}\,\dd q_1\,\dd\Omega_2,
\end{equation}
where $\int\dd\Omega_2=4\pi$. Hence
\begin{align}\label{eq:box_generic_threshold}
 &\lim_{x\to k_1}\frac{1}{\sqrt{x^2-k_1^2}}
 \lim_{M\to\infty}M^3
 \left[-\frac{\Disc_{X_1}\mathcal I_4(-x,M,M,M)}{2\pi i}\right]
=
 \frac{4\pi}{k_1^2}\int_0^{k_1}
 \frac{\sqrt{q_1(k_1-q_1)}\,\dd q_1}
 {|\vec K_2-q_1\hat n|\,|\vec K_3-q_1\hat n|}.
\end{align}
Both squared denominators in this one-dimensional integral are quadratic in $q_1$. At generic non-collinear spatial momenta, their four roots together with $0,k_1$ are distinct. The corresponding square root is therefore associated with a genus-two curve. If $|\vec K_3|\to\infty$ with $\vec K_3\cdot\hat n$ fixed, then
\begin{equation}\label{eq:box_generic_elliptic_limit}
 \lim_{|\vec K_3|\to\infty}|\vec K_3|
 \int_0^{k_1}
 \frac{\sqrt{q_1(k_1-q_1)}\,\dd q_1}
 {|\vec K_2-q_1\hat n|\,|\vec K_3-q_1\hat n|}
 =\int_0^{k_1}
 \frac{\sqrt{q_1(k_1-q_1)}\,\dd q_1}
 {|\vec K_2-q_1\hat n|}.
\end{equation}
For $\vec K_2$ not parallel to $\hat n$, the quartic under the square root has four distinct branch points whose cross-ratio varies with the external momenta. This gives a nontrivial elliptic period. The calculation already indicates that the complete correlator can not be expressed with multiple polylogarithms. The following example exhibits the elliptic function without taking this sequence
of large-energy and threshold limits.

\paragraph{Elliptic discontinuity of the full correlator.}
The preceding limit suggests that energy discontinuities can reveal features of the integrated box that are not evident from its basic integrands. The usual optical theorem motivates studying discontinuities because they replace propagators by on-shell constraints. Here we apply the same elementary idea to an energy denominator of the in-in correlator: $\Disc(1/(L+i0))=-2\pi i\,\delta(L)$. 

Consider the spatial configuration
\begin{equation}
 \vec k_1=\vec k_3=y\hat n,\quad
 \vec k_2=\vec k_4=-y\hat n,\quad
 q_1=q_3=|\vec q_1|,\quad
 q_2=q_4=|\vec q_1+y\hat n|,
\end{equation}
with $y>0$. Set $X_3=X_4=y$ and continue $X_1=X_2=-x\pm i0$, where $0<x<y$. In this region $q_1+q_2\geq y$ ensures that all tube denominators remain positive
except $L_{12}=X_1+X_2+q_1+q_3=2(q_1-x)$. Its discontinuity contributes $\tfrac12\delta(q_1-x)$. Among the 60 paired time orderings of the box, precisely 12 contain this tube: four with all vertices on one side of $t=0$, four with
three on one side, and four with two on each side. At $q_1=x$, the sum of these terms is
\begin{equation}\label{eq:box_residue_sum}
 \sum_{\text{12 time orderings}}
 \left.\prod_{L\ne L_{12}}\frac1L\right|_{q_1=x}
 =\frac{4y(2y+q_2)}
 {q_2(y^2-x^2)(q_2+y-x)(q_2+y+x)}.
\end{equation}
The product on the left contains the three tube denominators of each time ordering. In particular, the sum is taken before the remaining loop integration.

On the cut $q_1=x$, the angular measure in four dimensions is
\begin{equation}\label{eq:box_elliptic_curve}
 \dd\Omega_3=\frac{2\pi q_2Y(q_2)}{x^2y^2}\,\dd q_2,
 \quad
 Y(q_2)^2=[(y+x)^2-q_2^2][q_2^2-(y-x)^2],
~ y-x\leq q_2\leq y+x.
\end{equation}
Combining it with the factor $1/(8q_1q_2q_3q_4)$ in \eqref{eq:tubing_integral} and the half-residue of $L_{12}$ gives
\begin{equation}\label{eq:box_elliptic_cut}
 -\frac{\Disc_{12}\mathcal I_4}{2\pi i}
 =\frac{\pi}{2xy(y^2-x^2)}
 \int_{y-x}^{y+x}
 \frac{(2y+q_2)Y(q_2)\,\dd q_2}
 {q_2^2(q_2+y-x)(q_2+y+x)}.
\end{equation}
Here $\Disc_{12}$ means that $X_1$ and $X_2$ approach $-x$ from the same side of the cut, with $X_3,X_4$ held fixed. The identity
\begin{equation}
 \frac{(2y+q_2)Y(q_2)}
 {q_2^2(q_2+y-x)(q_2+y+x)}
 =\frac1{Y(q_2)}
 \left(
 -q_2+\frac{3y^2+x^2}{q_2}
 -\frac{2y(y^2-x^2)}{q_2^2}
 \right).
\end{equation}
reduces the integration to elementary terms and the complete elliptic integral of the second kind. Making the substitution $q_2^2=(y-x)^2+4xy\sin^2\theta$ and writing
$E(m)=\int_0^{\pi/2}\dd\theta\sqrt{1-m\sin^2\theta}$, the result is
\begin{equation}\label{eq:box_elliptic_result}
 -\frac{\Disc_{12}\mathcal I_4}{2\pi i}
 =\frac{\pi^2(x^2+y^2)}{2xy(y^2-x^2)^2}
 -\frac{\pi}{x(y-x)^2(y+x)}
 E\!\left(\frac{4xy}{(x+y)^2}\right).
\end{equation}
The four branch points of $Y(q_2)$ are $\pm(y-x)$ and $\pm(y+x)$. For $0<x<y$ they are distinct, and the argument of $E$ varies continuously between zero and one. The elliptic term has a nonzero coefficient in the complete time-ordering sum. Thus the answer is not, at generic kinematics, a finite combination of multiple polylogarithms with algebraic arguments and prefactors, despite the $d\log$ representation of its basic integrands \cite{Duhr:2020gdd}.

\paragraph{Discontinuities of the basic integrals.}
The preceding calculation establishes an elliptic contribution in the complete correlator. We now return to the Baikov representation and work at generic nondegenerate spatial kinematics to understand where this dependence enters the basic integrals. Consider $I_{1,2,3}$, whose three tube energies are
\begin{equation}
 z_a=A_a+q_1+q_{a+1},\qquad a=1,2,3,
\end{equation}
with $A_1=X_1$, $A_2=X_1+X_2$, and $A_3=X_1+X_2+X_3$, as in Section~\ref{sec:baikov}.

For this integral, the Baikov representations in $d=3$ and $d=4$ are
\begin{align}\label{eq:box_reduced_d4}
 \left.I_{1,2,3}\right|_{d=3}
 =\frac{\sqrt{\det H}}2
 \int_\Sigma
 \frac{\dd z_1\,\dd z_2\,\dd z_3}
 {z_1z_2z_3\sqrt{\Delta(z)}},\quad
 \left.I_{1,2,3}\right|_{d=4}
 =\frac{\pi}{4}
 \int_\Sigma
 \frac{\dd z_1\,\dd z_2\,\dd z_3}
 {z_1z_2z_3\sqrt{\Omega(z)}}.
\end{align}
Here $H$, $\Omega$, $\Delta$, $\Sigma$ are defined in Section~\ref{sec:baikov}, and $\left.I_{1,2,3}\right|_{d=3}$ is shown in Appendix~\ref{app:baikov_n_minus_1}.

We next take discontinuities in the tube energies. The $A_a$ are treated as independent variables, with the spatial momenta fixed. Each discontinuity is taken at fixed loop momentum before changing to the Baikov variables, so that
\begin{equation}
 -\frac{\Disc_{A_a}}{2\pi i}\frac1{z_a}
 =\delta(z_a).
\end{equation}
Although the domain $\Sigma$ depends on the $A_a$, the discontinuities therefore simply impose $z_a=0$ on the original momentum integral. Taking discontinuities in $A_1$ and $A_2$ leaves
\begin{align}\label{eq:box_double_cut_comparison}
 \frac{\Disc_{A_1}\Disc_{A_2}I_{1,2,3}}
 {(2\pi i)^2}
 =\begin{cases}
 \displaystyle
 \frac{\sqrt{\det H}}2
 \int\frac{\dd z_3}
 {z_3\sqrt{\Delta(0,0,z_3)}},
 &d=3,\\[10pt]
 \displaystyle
 \frac{\pi}4
 \int\frac{\dd z_3}
 {z_3\sqrt{\Omega(0,0,z_3)}},
 &d=4.
 \end{cases}
\end{align}
The integration ranges are determined by the physical domain $\Sigma$ restricted to $z_1=z_2=0$. The coefficients of $\Delta$ and $\Omega$ still depend on the external energies and spatial momenta.

In $d=3$, $\Delta(0,0,z_3)$ is generically a square-free quartic in $z_3$, so the remaining integral can be elliptic. In $d=4$, $\Omega(0,0,z_3)$ is quadratic, and the integral reduces to logarithms of algebraic functions. The endpoints are algebraic and do not introduce a further elliptic integration at this step.

The next integration in $d=4$ is different. A single tube energy discontinuity is
\begin{equation}\label{eq:box_basic_single_cut}
 -\frac{\Disc_{A_1}I_{1,2,3}}{2\pi i}
 \bigg|_{d=4}
 =\frac18\int\frac{\dd^4\vec q_1\,
 \delta(A_1+q_1+q_2)}
 {q_1q_2q_3q_4
 (A_2+q_1+q_3)(A_3+q_1+q_4)}.
\end{equation}
Multiplying by $A_2A_3$ and sending both to infinity, with $A_3>A_2>0$, gives
one eighth of the phase-space integral in \eqref{eq:box_large_energy_cut}, with $x=-A_1$. Its threshold coefficient is therefore one eighth of \eqref{eq:box_generic_threshold}, and the elliptic limit \eqref{eq:box_generic_elliptic_limit} is inherited by
this single cut. Thus an individual basic box integral can already have a non-polylogarithmic single tube discontinuity in $d=4$, even though its double tube discontinuity is logarithmic. For $d=3$, the corresponding elliptic dependence can appear one integration earlier, in the double discontinuity \eqref{eq:box_double_cut_comparison}. The comparison for different numbers of tube-energy discontinuities is summarized in Table~\ref{tab:box_discontinuities}.

\begin{table}[htbp]
 \centering
 \begin{tabular}{@{}ccc@{}}
 \toprule
 Tube discontinuities & $d=3$ & $d=4$\\
 \midrule
 Single & Double integral & Double integral\\
 Double & Elliptic integral & Logarithmic integral\\
 Triple & Algebraic & Algebraic\\
 \bottomrule
 \end{tabular}
 \caption{Analytic structure of successive tube discontinuities of the basic box integral $I_{1,2,3}$.}
 \label{tab:box_discontinuities}
\end{table}

The algebraic threefold tube residue is consistent with the $d\log$ representation at $d=4$. It does not constrain the analytic structure of the remaining integrations: although $\Delta$ is absent from the $d=4$ integrand, the physical domain is still restricted by $\Delta\geq0$. This explains why the local leading singularities can be algebraic even when the integrated box contains elliptic
functions.

\paragraph{Comparison with dressed cuts.}
We briefly compare our tube-energy cuts with the cosmological dressing rules~\cite{Chowdhury:2025ohm,Chowdhury:2026upp},
where discontinuities in squared external energies and momentum invariants are related to cuts of auxiliary propagators and flat-space Feynman integrals, respectively.
Fourier transforming the time-dependent factors in
\eqref{eq:kernel}, the dressed box takes the form
\begin{equation}
 \mathcal I_4=
 \int\dd^4\vec\ell
 \int\prod_{i=1}^4\frac{\dd\omega_i}{2\pi}
 \prod_{i=1}^4
 \frac{1}{\omega_i^2+q_i^2}
 \frac{2X_i}{(\omega_{i+1}-\omega_i)^2+X_i^2},
 \qquad \omega_5=\omega_1.
\end{equation}
The two factors correspond to the ordinary and auxiliary
propagators. Consider a generalized cut putting all four
ordinary propagators and the first three auxiliary propagators
on shell. On the branch
$(\omega_1,\omega_2,\omega_3,\omega_4)
=(iq_1,-iq_2,-iq_3,-iq_4)$,
the auxiliary poles can be chosen as
\begin{equation}
 \begin{aligned}
 X_1+q_1+q_2=z_1=0,\quad
 X_2+q_3-q_2=z_2-z_1=0,\quad
 X_3+q_4-q_3=z_3-z_2=0.
 \end{aligned}
\end{equation}
Their common solution is precisely the maximal tube-pole
configuration $z_1=z_2=z_3=0$ of $I_{1,2,3}$,
and the change between these linear constraints has unit
Jacobian. The fourth auxiliary propagator remains uncut and gives exactly the coefficient of $I_{1,2,3}$
in \eqref{eq:reduction}. Combining the two globally reversed
ordinary-energy branches reproduces the measure
$\dd^4\vec\ell/(8q_1q_2q_3q_4)$.
The remaining threefold tube cut is then evaluated using
\eqref{eq:reduced}, giving
\begin{equation}
 \left.\mathcal I_4\right|_{\mathrm{selected\ cut}}
 =
 \frac{2X_4}{X_4^2-(X_1+X_2+X_3)^2}
 \frac{\pi}{4\sqrt{\Omega(0)}}.
\end{equation}
Thus the selected dressed cut reproduces both our reduction
coefficient and the algebraic tube leading singularity.
Note that this is a correspondence between local cut contributions, rather than an equality with the full sequential discontinuities in squared kinematic variables, which require the appropriate energy-branch sums.

\section{Discussion}\label{sec:discussion}

We have studied the analytic structure of scalar
one-loop $n$-gon correlators by combining an
integrand-level reduction with a Baikov representation
of the spatial loop integral. The reduction expresses
the full correlator in terms of $n\,n!/2$ basic-integral
terms, organized into $n2^{n-2}$ groups. At $d=n$, each basic integral admits a square-root representation whose local leading singularities can be normalized to unity, together with an explicit $d\log$ form. The bubble and triangle have hyperbolic-volume interpretations, consistent with
the known dilogarithmic triangle result~\cite{Pimentel:2026kqc,Henn:2026lfz}.
For the complete box at $d=4$, we have evaluated
an external-energy discontinuity in closed form
and found a nonvanishing elliptic contribution
after summing all time orderings.

Our result is closely related to the recent analysis
of McCune and Schwartz~\cite{McCune:2026qfy}.
They showed that although the master integrals of
the one-loop triangle wavefunction contain elliptic
sectors, their periods cancel in the physical
coefficient, which is dilogarithmic. For the box,
they instead found elliptic threshold discontinuities
and, in special kinematics, contributions involving
periods of a K3 surface. Their analysis also extends to correlators,
including elliptic contributions to the box
in three spatial dimensions.
Our explicit $d=n=4$ discontinuity provides
a complementary result: an elliptic contribution
survives the complete time-ordering sum
despite algebraic local tube leading
singularities.
Taken together, these results emphasize
that the function space of an integrated
observable cannot be inferred from its
local residues alone~\cite{Duhr:2020gdd,Benincasa:2024ptf}.

A natural next step is to determine the box at
generic external kinematics. In this respect,
the recent work of Sachs~\cite{Sachs:2026gcr}
provides a particularly useful complementary
approach. Starting from dressed flat-space
amplitudes, Sachs developed differential equations
and IBP reductions for in-in correlators, exploiting
partial-fraction identities among the auxiliary
energy propagators. For the cosmological triangle,
the flat-space leading singularity already controls
the relevant top sector of the differential system
and explains the dilogarithmic structure of the
answer. Our Baikov representation instead makes
the algebraic leading singularities and integration
domain explicit before constructing a differential
system. Combining these approaches could provide
an efficient route to differential equations for
the basic box integrals and to a characterization
of their non-polylogarithmic sectors~\cite{Henn:2013pwa,Herrmann:2019upk,
Arkani-Hamed:2023kig,He:2024olr,
Baumann:2024mvm,Chen:2023iix,De:2023xue}.

An alternative route is to reconstruct the
correlator from its discontinuities. Our comparison
with the dressing rules identifies a correspondence
between selected dressed cuts and local tube-pole
contributions, but a systematic relation between
linear tube-energy cuts and discontinuities in
squared kinematic variables remains to be established.
Such a relation, together with suitable dispersion
relations and boundary conditions, may constrain
the full answer without evaluating every basic
integral separately~\cite{Goodhew:2021oqg,
Chowdhury:2025ohm,Chowdhury:2026upp,
Colipi-Marchant:2025oin,Das:2025qsh}.

The integration domain is equally important.
Although $\Delta(z)$ drops out of the Baikov
integrand at $d=n$, it continues to determine
the physical boundaries. Understanding how these
boundaries enter the analytic continuation may
help explain the transition from the hyperbolic
geometry of the low-point integrals~\cite{Davydychev:1997wa,Schnetz:2010pd,
Bourjaily:2019dix}
to the elliptic structures of the box.
The generalized Euler-integral representation
of cosmological correlators and its Landau
analysis~\cite{Katsevich:2026kgc} offer a
complementary perspective on this question.

Finally, the reduction itself suggests a
combinatorial comparison with graph-level
organizations of scattering amplitudes,
including causal representations of loop
integrals and differential-operator projections
onto individual cubic diagrams~\cite{Capatti:2020ytd,Kromin:2022txz,Dong:2025lrf}.
Related in-in constructions based on
zonotopes, stringy integrals and the simplified
Born rule~\cite{He:2026rpw,Zhu:2026xil}
may help identify relations among the basic
integrals before spatial integration. The collapsed and ultraviolet
limits derived in Appendix~\ref{app:limits}
provide further checks of their relation
to flat-space amplitudes~\cite{Qin:2023bjk,Xianyu:2022jwk,
Benincasa:2024lxe,Beneke:2023wmt}.

\acknowledgments
We thank Song He and Chandramouli Chowdhury for helpful discussions and early collaborations on this work. The authors also thank Xuhang Jiang for insightful discussions. OpenAI's ChatGPT-6 Astra has been used for preparing graphs, setting up the manuscript and verifying the derivations. The authors have reviewed and verified all AI-assisted material and scientific
conclusions and take full responsibility for the content. This work is supported in part by National Natural Science Foundation of China under Grant No.12225510 and No.12247103, and by the New Cornerstone Science Foundation.
\appendix
\section{Details of Baikov representation}\label{app:details_Baikov}
\subsection{Definitions of polynomials and integral domain}
 We let $\vec K_a=\sum_{i=1}^{a}\vec k_i$ ($a=1,\ldots,n-1$), and generally these $n-1$ vectors are linearly independent. The corresponding Gram matrix and the Baikov polynomial appeared in \eqref{eq:change_of_variable}
are respectively given as
\begin{align}\label{eq:Gram_H}
 &H_{ab}=\vec K_a\cdot\vec K_b,\qquad \det H>0,\\
 &\mathcal B=\det\operatorname{Gram}
 (\vec q_1,\vec K_1,\ldots,\vec K_{n-1})=\det H\left[q_1^2-
 \sum_{a,b=1}^{n-1}(\vec q_1\cdot\vec K_a)
 (H^{-1})_{ab}(\vec q_1\cdot\vec K_b)\right].
 \label{eq:Baikov_polynomial}
\end{align}
After the change of variables and define $C_{ij}=c_{i,j+1}$, it is allowed to represent the edge energies $q$'s in terms of the tube variables $z$'s and $q_1$ as
\begin{equation}\label{eq:y_sigma}
q_{i+1}=y_i(z)+\sigma_iq_1\implies\begin{cases}
     y_i(z)=\displaystyle\sum_{j=1}^{n-1}(C^{-1})_{ij}(z_j-A_j),\\
     \sigma_i=-\displaystyle\sum_{j=1}^{n-1}(C^{-1})_{ij}c_{j1}.
 \end{cases}
\end{equation}
Here $i=1,\ldots,n-1$, and one can further show that
\begin{equation}
      \sigma_i=\pm1.
\end{equation} 
 Therefore the
Baikov polynomial is quadratic in the remaining variable $q_1$
\begin{equation}\label{eq:quadratic_equation_q1}
    \mathcal B(q_1,z)=-\Omega(z)q_1^2+\beta(z)q_1+\gamma(z),
\end{equation}
where
\begin{eqn}\label{eq:quadratic_baikov}
 \Omega(z)&=\sum_{a,b=1}^{n-1}\sigma_a\sigma_b
 y_a(z)(\operatorname{adj}H)_{ab}y_b(z)-\det H,\\
 \beta(z)&=-\sum_{a,b=1}^{n-1}
 \bigl(y_a(z)^2-H_{aa}\bigr)(\operatorname{adj}H)_{ab}
 \sigma_b y_b(z),\\
 \gamma(z)&=-\frac14\sum_{a,b=1}^{n-1}
 \bigl(y_a(z)^2-H_{aa}\bigr)(\operatorname{adj}H)_{ab}
 \bigl(y_b(z)^2-H_{bb}\bigr),
\end{eqn}
and the adjoint of matrix $H$ can be computed out of its inverse and determinant --- $\operatorname{adj}H=H^{-1}\det H $.
The domain is given by the requirement of letting all the edge energies $q_i$ and the Baikov polynomial to be positive
\begin{equation}
    \Sigma=\Bigl\{z\Big|\  q_1>0,\quad\mathcal B(q_1,z)>0,\quad y_i(z)+\sigma_iq_1>0\quad(i=1,\ldots,n-1)\Bigr\}.
\end{equation}
After integrate out $q_1$, it becomes equivalently the restriction of letting the quadratic equation \eqref{eq:quadratic_equation_q1} have real solutions (which is to say, let the discriminant $\Delta$ positive) and other edge energies positive
\begin{equation}
    \Sigma=\left\{z|\Delta(z)>0,\quad y_i(z)+\sigma_iq_1>0\quad(i=1,\ldots,n-1)\right\},
\end{equation}
with the discriminant
\begin{equation}
    \Delta(z)=\beta(z)^2+4\Omega(z)\gamma(z).
\end{equation}

\subsection{Baikov representation at
\texorpdfstring{$d=n-1$}{d=n-1}}
\label{app:baikov_n_minus_1}

Although the Baikov representation in
Section~\ref{sec:baikov} was derived for $d\geq n$,
the integrated formula \eqref{eq:integrated_baikov}
admits an extension to $d=n-1$. We first obtain
this result by dimensional continuation and then
derive it independently from the loop-momentum
measure in $n-1$ dimensions. 

\paragraph{Dimensional continuation.}
Setting $d=n-1+\epsilon$ in
\eqref{eq:integrated_baikov} gives
\begin{align}
I^{(n)}
={}&
\frac{2^{2-n}\pi^{\epsilon/2}
\Gamma(\epsilon/2)}
{\Gamma(\epsilon)}
(\det H)^{(1-\epsilon)/2}
\int_\Sigma
\frac{\dd z_1\cdots\dd z_{n-1}}
{z_1\cdots z_{n-1}}\,
\frac{\Delta(z)^{(-1+\epsilon)/2}}
{\Omega(z)^{\epsilon/2}}.
\label{eq:baikov_dimensional_limit}
\end{align}
Although both Gamma functions diverge as
$\epsilon\to0^+$, their ratio has a finite limit
\begin{equation}
\lim_{\epsilon\to0^+}
\frac{\Gamma(\epsilon/2)}{\Gamma(\epsilon)}
=2.
\end{equation}
Moreover, for fixed $z$ in the regular interior
of the physical integration domain
\begin{equation}
\lim_{\epsilon\to0^+}
\frac{\Delta(z)^{(-1+\epsilon)/2}}
{\Omega(z)^{\epsilon/2}}
=\frac{1}{\sqrt{\Delta(z)}}.
\end{equation}
Provided the dimensional limit can be taken under
the integral, we therefore obtain
\begin{equation}
\left.I^{(n)}\right|_{d=n-1}
=
2^{3-n}\sqrt{\det H}
\int_\Sigma
\frac{\dd z_1\cdots\dd z_{n-1}}
{z_1\cdots z_{n-1}\sqrt{\Delta(z)}}.
\label{eq:baikov_n_minus_1}
\end{equation}

\paragraph{Direct derivation at $d=n-1$.}
We now establish \eqref{eq:baikov_n_minus_1}
without dimensional continuation.
When $d=n-1$, the external vectors
$\vec K_1,\ldots,\vec K_{n-1}$ span the full
spatial momentum space. Consequently, the
Baikov polynomial \eqref{eq:Baikov_polynomial}
vanishes identically on physical loop momenta
\begin{equation}
\mathcal B
=
\det\operatorname{Gram}
(\vec\ell,\vec K_1,\ldots,\vec K_{n-1})
=0.
\end{equation}
The physical integration measure is therefore
supported on the hypersurface $\mathcal B=0$,
rather than on a region with $\mathcal B>0$.
This gives the distributional measure
\begin{equation}
\dd^{n-1}\vec\ell
=
2\sqrt{\det H}
\prod_{i=1}^{n}q_i\,\dd q_i\,
\delta(\mathcal B).
\label{eq:baikov_delta_measure}
\end{equation}
The equality is understood under integration
over the positive internal energies $q_i>0$.

Substituting \eqref{eq:baikov_delta_measure}
into the definition \eqref{eq:z_variables},
and replacing $q_2,\ldots,q_n$ by the tube
energies $z_1,\ldots,z_{n-1}$, gives
\begin{align}
\left.I^{(n)}\right|_{d=n-1}
={}&
2^{2-n}\sqrt{\det H}
\int_{\substack{
q_1>0,\;
y_i(z)+\sigma_iq_1>0}}
\frac{\dd q_1\,\dd z_1\cdots\dd z_{n-1}}
{z_1\cdots z_{n-1}}
\delta\bigl(
-\Omega(z)q_1^2+\beta(z)q_1+\gamma(z)
\bigr).
\label{eq:baikov_delta_representation}
\end{align}
At fixed $z$, the quadratic equation
$\mathcal B(q_1,z)=0$ has the roots
\begin{equation}
q_1^\pm
=
\frac{\beta(z)\pm\sqrt{\Delta(z)}}
{2\Omega(z)},
\end{equation}
and they contribute equally, yielding
\begin{equation}
\int\dd q_1\,
\delta\bigl(
-\Omega(z)q_1^2+\beta(z)q_1+\gamma(z)
\bigr)
=
\frac{2}{\sqrt{\Delta(z)}}.
\end{equation}
Substituting this into
\eqref{eq:baikov_delta_representation}
reproduces \eqref{eq:baikov_n_minus_1},
including its overall normalization.

\section{Hyperbolic Volume Interpretation of Bubbles and Triangles}
\subsection{The bubble at \texorpdfstring{$d=n=2$}{d=n=2}}
\label{sec:bubble_geometry}
\begin{figure}[!htbp]
 \centering
 \hspace*{4em}\includegraphics[width=.94\textwidth]{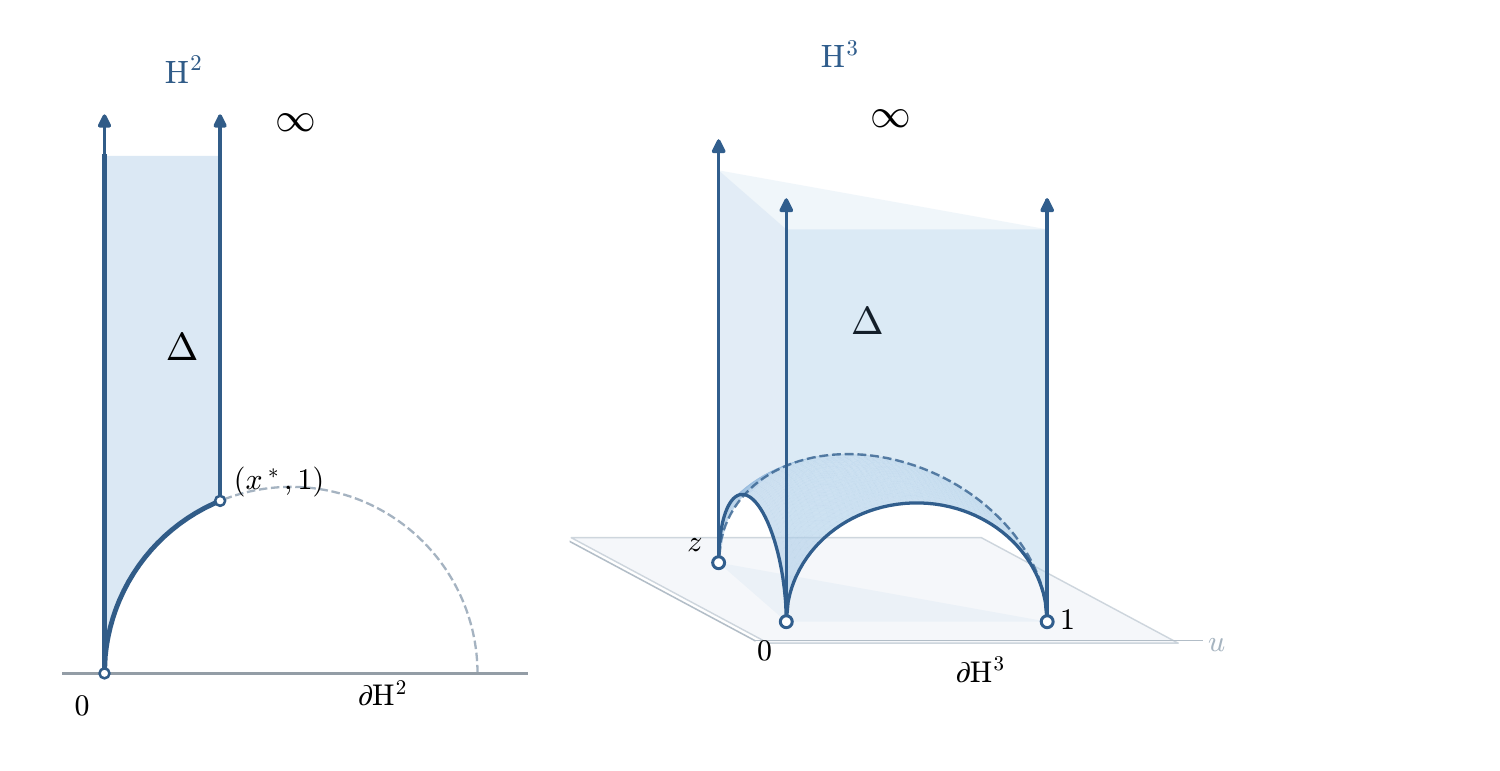}
 \caption{Left: the bubble's hyperbolic triangle for $0<A<k$.
 The shaded region lies above the geodesic arc and extends to the ideal
 vertex at infinity. Right: the ideal tetrahedron
in $\mathbb H^3$, the building block of the triangle's
four-term volume representation.}
 \label{fig:low_point_geometry}
\end{figure}
We begin with the bubble, for which the basic integral
can be evaluated in arbitrary spatial dimension $d$.
Let $k=|\vec k_1|=|\vec k_2|>0$, and let $A=X_1$ or
$X_2$ denote the vertex energy of either basic integral.
The Baikov representation \eqref{eq:integrated_baikov}
simplifies considerably for $n=2$.
In this case
\begin{equation}
    I^{(2)}
    =\frac{2^{2-d}\pi^{d/2}}{\Gamma(d/2)}
    \int_{A+k}^{\infty}
    \frac{\bigl[(z-A)^2-k^2\bigr]^{(d-3)/2}}
    {z}\,\dd z.
    \label{eq:bubble_general_d}
\end{equation}
The result can be expressed in terms of a
Gauss hypergeometric function
\begin{align}
    I^{(2)}
    ={}&
    \frac{\pi^{d/2}k^{d-2}}
    {(A+k)\Gamma(d/2)}
    \frac{
    \Gamma\bigl(\frac{d-1}{2}\bigr)\Gamma(3-d)}
    {\Gamma\bigl(\frac{5-d}{2}\bigr)}
    {}_2F_1\left(
    1,\frac{d-1}{2};
    \frac{5-d}{2};
    \frac{A-k}{A+k}
    \right).
    \label{eq:bubble_hypergeometric}
\end{align}
Specializing to $d=n=2$ gives
\begin{equation}
    {
    \frac{1}{\pi}I^{(2)}\big|_{d=2}
    =
    \begin{cases}
    \displaystyle
    \frac{\arccos(A/k)}{\sqrt{k^2-A^2}},
    &0<A<k,\\[8pt]
    \displaystyle
    \frac1k,
    &A=k,\\[8pt]
    \displaystyle
    \frac{\operatorname{arcosh}(A/k)}
    {\sqrt{A^2-k^2}},
    &A>k.
    \end{cases}
    }
    \label{eq:bubble_log}
\end{equation}
Here $\operatorname{arcosh}(A/k)
=\log[(A+\sqrt{A^2-k^2})/k]$ for $A>k$.
Since $\Omega(0)=A^2-k^2$, this expression
also exhibits the square-root factor
determined by the local leading singularity
in \eqref{eq:local_LS}.

We now give a hyperbolic-volume interpretation.
For the bubble, consider $0<A<k$.
In the upper-half-plane model of $\mathbb H^2$,
with the boundaries
\begin{equation}
    y^2=\frac{2k}{\sqrt{k^2-A^2}}x-x^2, \qquad x=0,\qquad x=x^*,
\end{equation}
and together with vertices (see Figure \ref{fig:low_point_geometry})
\begin{equation}
    {
    A_1=(0,0),\qquad
    A_2=\left(\sqrt{\frac{k-A}{k+A}},1\right)=(x^*,1),
    \qquad A_3=\infty,
    }
    \label{eq:bubble_vertices}
\end{equation}
the volume envoloped by the hyperbolic triangle is
\begin{align}
    \operatorname{Area}_{\mathbb H^2}
    \Delta(A_1,A_2,A_3)
    =\arccos\frac Ak=\frac{\sqrt{k^2-A^2}}{\pi}
    I^{(2)}\big|_{d=2}
    \label{eq:bubble_area}
\end{align}
Analytic continuation from $0<A<k$ to
$A>k$ through the lower half-plane sends
$\sqrt{k^2-A^2}\to \mathrm i\sqrt{A^2-k^2}$
and $\arccos(A/k)\to \mathrm i\operatorname{arcosh}(A/k)$.
Thus the normalized basic bubble integral
admits an interpretation as an analytically
continued hyperbolic triangle area, in close
analogy with the geometry of ordinary
one-loop Feynman integrals~\cite{Davydychev:1997wa,Schnetz:2010pd,
Bourjaily:2019dix}.

\subsection{The triangle at \texorpdfstring{$d=n=3$}{d=n=3}}
\label{sec:triangle}

The hyperbolic area of the bubble has a three-dimensional analogue for
basic triangle integrals. Consider $I_{3,2}$ and $I_{3;1}$ in
\eqref{eq:triangle_reduction}. With $U=q_1+q_3$ and $V=q_1+q_2$, their
tube denominators are $U+A_1$ and $V+A_2$, where
$(A_1,A_2)=(X_3,X_2+X_3)$ and $(X_3,X_1)$, respectively.
At $d=3$, \eqref{eq:reduced} gives
\begin{align}\label{eq:triangle_geometric_integral}
 I^{(3)}\big|_{d=3}
 &=\frac{\pi}{2}\int_{\substack{U\ge |\vec k_3|,\;V\ge |\vec k_1|\\
                          |U-V|\le |\vec k_2|}}
 \frac{\dd U\,\dd V}{(U+A_1)(V+A_2)\sqrt{\Omega(U,V)}},\notag\\
 \Omega(U,V)&=H_{11}U^2-2H_{12}UV+H_{22}V^2-\det H.
\end{align}
Here $H$ is the Gram matrix of \eqref{eq:Gram_H}, with
$H_{11}=|\vec k_1|^2$ and $H_{22}=|\vec k_3|^2$.
The radicand at the tube poles is
$\Omega(0)=\Omega(-A_1,-A_2)$, as in \eqref{eq:LS_root}.
The integration region has three finite corners
$(|\vec k_3|,|\vec k_1|)$,
$(|\vec k_1|+|\vec k_2|,|\vec k_1|)$ and
$(|\vec k_3|,|\vec k_2|+|\vec k_3|)$, together with one end at infinity.

To see why a three-dimensional volume appears, consider first a single
ideal tetrahedron in the upper-half-space model of $\mathbb H^3$
\begin{equation}\label{eq:triangle_h3_metric}
 \dd s^2=\frac{\dd r^2+\dd w\,\dd\bar w}{r^2},
 ~ r>0,\qquad
 \dd\operatorname{Vol}_{\mathbb H^3}=\frac{\dd^2w\,\dd r}{r^3}.
\end{equation}
By an isometry its four vertices can be placed at
$0,1,z,\infty$ on the conformal boundary, with $\operatorname{Im}z>0$.
The three faces meeting at infinity are vertical geodesic planes above
the sides of the Euclidean triangle with vertices $0,1,z$, as shown in Figure~\ref{fig:low_point_geometry}. The remaining face is the hemisphere
$r^2+|w-c|^2=R^2$, where $c$ and $R$ are the circumcentre and
circumradius of that boundary triangle (so $R=|c|$).
Consequently, at fixed $w$ inside the boundary triangle, the tetrahedron
extends from $r=\sqrt{R^2-|w-c|^2}$ to $r=\infty$. Integrating
in the vertical direction gives
\begin{align}\label{eq:triangle_h3_volume}
 \operatorname{Vol}_{\mathbb H^3}\Delta(0,1,z,\infty)
 &=\int_{\operatorname{conv}\{0,1,z\}}\dd^2 w
   \int_{\sqrt{R^2-|w-c|^2}}^\infty\frac{\dd r}{r^3}=\frac12\int_{\operatorname{conv}\{0,1,z\}}
   \frac{\dd^2 w}{R^2-|w-c|^2}.
\end{align}
Evaluating this integral gives the Bloch--Wigner volume
\begin{equation}\label{eq:triangle_h3_bloch}
 \operatorname{Vol}_{\mathbb H^3}\Delta(0,1,z,\infty)
 =\operatorname{Im}\operatorname{Li}_2(z)
  +\log|z|\,\arg(1-z).
\end{equation}
On the branch where $\bar z$ is the complex conjugate of $z$, the
same result can be written as
\begin{equation}\label{eq:triangle_two_sheet_volume}
 2\mathrm i\operatorname{Vol}_{\mathbb H^3}\Delta(0,1,z,\infty)
 =\operatorname{Li}_2(z)-\operatorname{Li}_2(\bar z)
 +\frac12\log(z\bar z)\log\frac{1-z}{1-\bar z}.
\end{equation}
This form is convenient for analytic continuation away from conjugate
arguments.

We now return to \eqref{eq:triangle_geometric_integral}. The two tangents
to the conic $\Omega(U,V)=0$ through the intersection of the tube poles,
$(U,V)=(-A_1,-A_2)$, have slopes
\begin{equation}\label{eq:triangle_tangent_slopes}
 C_\pm=\frac{A_1A_2-H_{12}\pm\sqrt{\Omega(0)}}{A_2^2-H_{11}}.
\end{equation}
Their discriminant is $4\Omega(0)$: the same square root that
normalizes the tube leading singularity also enters the tetrahedral
shape parameters. To make the connection explicit, set
\begin{equation}\label{eq:triangle_conic_rationalization}
 \xi=\frac{\sqrt{\Omega(U,V)}-|\vec k_3|V+H_{12}}
 {U-|\vec k_3|},\qquad
 \tau=\frac{\sqrt{\Omega(U,V)}+|\vec k_3|V-H_{12}}
 {U-|\vec k_3|}.
\end{equation}
This maps the integration domain to the rectangle
$|\vec k_2|-|\vec k_3|\le\xi\le|\vec k_1|$ and
$|\vec k_2|+|\vec k_3|\le\tau<\infty$.
On the conjugate-root branch, the chain rule applied to
\eqref{eq:triangle_two_sheet_volume} gives
\begin{align}\label{eq:triangle_volume_derivative}
 &\frac{\partial^2}{\partial\xi\,\partial\tau}
 \left[2i\,\operatorname{Vol}_{\mathbb H^3}\Delta\!\left(
 0,1,\frac{U+A_1}{(V+A_2)C_+},\infty\right)\right]=\frac{\sqrt{\Omega(0)}}
 {(U+A_1)(V+A_2)\sqrt{\Omega(U,V)}}
 \left|\frac{\partial(U,V)}{\partial(\xi,\tau)}\right|.
\end{align}
Thus the Baikov measure is the mixed derivative of a hyperbolic-volume
function. Integration over the rectangle leaves an alternating sum of
its four corner values, with signs $+,-,-,+$. In particular,
$(U+A_1)/(V+A_2)$ takes the values
\begin{equation}
 \frac{A_1+|\vec k_3|}{A_2+|\vec k_1|},\quad
 \frac{A_1+|\vec k_1|+|\vec k_2|}{A_2+|\vec k_1|},\quad
 \frac{A_1+|\vec k_3|}{A_2+|\vec k_2|+|\vec k_3|},\quad 1.
\end{equation}
We obtain the explicit volume representation
\begin{align}\label{eq:triangle_explicit_volumes}
 I^{(3)}\big|_{d=3}
 =\frac{i\pi}{\sqrt{\Omega(0)}}\Bigg[\,&
 \operatorname{Vol}_{\mathbb H^3}\Delta\!\left(
 0,1,\frac{A_1+|\vec k_3|}{(A_2+|\vec k_1|)C_+},\infty\right)-\operatorname{Vol}_{\mathbb H^3}\Delta\!\left(
 0,1,\frac{A_1+|\vec k_1|+|\vec k_2|}
 {(A_2+|\vec k_1|)C_+},\infty\right)\notag\\
 &-\operatorname{Vol}_{\mathbb H^3}\Delta\!\left(
 0,1,\frac{A_1+|\vec k_3|}
 {(A_2+|\vec k_2|+|\vec k_3|)C_+},\infty\right)+\operatorname{Vol}_{\mathbb H^3}\Delta\!\left(
 0,1,\frac1{C_+},\infty\right)\Bigg]_{\rm cont}.
\end{align}
The equality is first understood on a conjugate-root
branch, where $C_-=C_+^*$ and the terms are ordinary
oriented hyperbolic volumes. The notation
$[\cdots]_{\rm cont}$ denotes analytic continuation
of the complete two-sheet expression, with both
$C_\pm$ roots and their logarithmic branches
continued together.

The two assignments of $(A_1,A_2)$ given above, together with their
permutations and the coefficients in \eqref{eq:triangle_reduction},
reconstruct the complete triangle correlator. The resulting algebraic
prefactors and dilogarithms agree with~\cite{Henn:2026lfz}.

\section{Routing templates and dihedral relabelings}\label{app:routing}
We define the rotation
\begin{equation}
    R^m:\quad(q_1,\dots,q_n)\to(q_{m+1},\dots,q_n,q_1,\dots,q_m)
\end{equation}
together with the reflection
\begin{equation}
    F_m:\quad(q_1,\dots,q_n)\to(q_m,q_{m-1},\dots,q_1,q_n,\dots,q_{m+1}).
\end{equation}
Note that $R^m$ could be achieved by doing the first transformation in $\vec k's$
\begin{equation}
    (\vec k_1,\dots,\vec k_n)\to(\vec k_{m+1},\dots,\vec k_n,\vec k_1,\dots,\vec k_m),
\end{equation}  
and then the following change of variables
\begin{equation}
    \vec\ell\to\vec\ell'=\vec\ell+\vec k_1+\cdots+\vec k_m.
\end{equation}
For $F_m$, they could be made by taking 
\begin{equation}
    (\vec k_1,\dots,\vec k_n)\to(\vec k_{m-1},\dots,\vec k_1,\vec k_n,\dots,\vec k_m)
\end{equation}
and the change of variables
\begin{equation}
    \vec\ell\to\vec\ell'=-\vec \ell-\vec k_1-\cdots-\vec k_{m-1}.
\end{equation}
Note that for both different ways of variation of loop momentum, the corresponding Jacobian is always 1, which will not affect the integral result. As the summary, the effect of dihedral permutations on $\vec k$'s are
\begin{equation}
    \begin{aligned}
        R^m:\quad&(\vec k_1,\dots,\vec k_n)\to(\vec k_{m+1},\dots,\vec k_n,\vec k_1,\dots,\vec k_m),\\
        F_m:\quad&(\vec k_1,\dots,\vec k_n)\to(\vec k_{m-1},\dots,\vec k_1,\vec k_n,\dots,\vec k_m).
    \end{aligned}
\end{equation}
\begin{quote}
    \it To find the independent integral under the dihedral group $D_n$, we could start with a set of edge energies, permuting q's with $D_n$ (also corresponding with the change of k's), and see the orbit of this group action along with all possible consistent vertex energies. 
\end{quote}
For example, at $n=2$ there is only one independent integral in under $D_2$ and we could choose its representative as $I_1^{(2)}$

\begingroup\scriptsize\SetTblrTemplate{head}{empty}
\begin{longtblr}{ccccc}
    \hline[1pt]
$D_2$ & $(\vec k_1',\vec k_2')$ &Edge Energies&Vertex Energies&Integrals\\\hline
$\mathbf1$&$(\vec k_1,\vec k_2)$&\SetCell[r=2]{c}$q_1+q_2$&\SetCell[r=2]{c}$X_1,X_2$&\SetCell[r=2]{c}$I^{(2)}_1$, $I_2^{(2)}$\\
$R^1$&$(\vec k_2,\vec k_1)$&&&\\\hline[1pt]
\end{longtblr}
\endgroup
At $n=3$, there is still only 1 representative $I_{1,2}^{(3)}$ (with the repetitive integral written in gray)
\begingroup\scriptsize\SetTblrTemplate{head}{empty}
\begin{longtblr}{ccccc}
    \hline[1pt]
    $D_3$ & $(\vec k_1',\vec k_2',\vec k_3')$ &Edge Energies&Vertex Energies&Integrals\\\hline
    \SetCell[r=2]{c}$\mathbf{1}$&\SetCell[r=2]{c}$(\vec k_1,\vec k_2,\vec k_3)$&\SetCell[r=2]{c}$(q_1+q_2,q_1+q_3)$&$(X_1,X_1+X_2)$&$I_{1,2}^{(3)}$\\
    &&&$(X_1,X_3)$&$I_{1;3}^{(3)}$\\
    \SetCell[r=2]{c}$R^1$&\SetCell[r=2]{c}$(\vec k_2,\vec k_3,\vec k_1)$&\SetCell[r=2]{c}$(q_2+q_3,q_2+q_1)$&$(X_2,X_2+X_3)$&$I_{2,3}^{(3)}$\\
    &&&$(X_2,X_1)$&$I_{2;1}^{(3)}$\\
    \SetCell[r=2]{c}$R^2$&\SetCell[r=2]{c}$(\vec k_3,\vec k_1,\vec k_2)$&\SetCell[r=2]{c}$(q_3+q_1,q_3+q_2)$&$(X_3,X_1+X_3)$&$I_{3,1}^{(3)}$\\
    &&&$(X_3,X_2)$&$I_{3;2}^{(3)}$\\
    \SetCell[r=2]{c}$F_1$&\SetCell[r=2]{c}$(\vec k_3,\vec k_2,\vec k_1)$&\SetCell[r=2]{c}$(q_1+q_3,q_2+q_1)$&$(X_3,X_2+X_3)$&$I_{3,2}^{(3)}$\\
    &&&$(X_3,X_1)$&$\color{gray}I_{3;1}^{(3)}$\\
    \SetCell[r=2]{c}$F_2$&\SetCell[r=2]{c}$(\vec k_1,\vec k_3,\vec k_2)$&\SetCell[r=2]{c}$(q_1+q_2,q_2+q_3)$&$(X_1,X_1+X_3)$&$I_{1,3}^{(3)}$\\
    &&&$(X_1,X_2)$&$\color{gray}I_{1;2}^{(3)}$\\
    \SetCell[r=2]{c}$F_3$&\SetCell[r=2]{c}$(\vec k_2,\vec k_1,\vec k_3)$&\SetCell[r=2]{c}$(q_3+q_2,q_1+q_3)$&$(X_2,X_1+X_2)$&$I_{2,1}^{(3)}$\\
    &&&$(X_2,X_3)$&$\color{gray}I_{2;3}^{(3)}$\\\hline[1pt]
\end{longtblr}
\endgroup
\subsection{Box routing classes}
However, at $n=4$, non-triviality emerges, in which there are three inequivalent classes under the dihedral action, represented by $I_{1,2,3}^{(4)}$, $I_{2,1,3}^{(4)}$ and $I_{1,3,2}^{(4)}$.

\noindent\textbf{Routing class 1.}

\begingroup\scriptsize\SetTblrTemplate{head}{empty}
\begin{longtblr}{ccccc}
\hline[1pt]
$D_4$ & $(\vec k_1',\vec k_2',\vec k_3',\vec k_4')$ & Edge Energies & Vertex Energies & Integrals\\
\hline
\SetCell[r=2]{c}$\mathbf 1$
&\SetCell[r=2]{c}$(\vec k_1,\vec k_2,\vec k_3,\vec k_4)$
&\SetCell[r=2]{c}$(q_1+q_2,q_1+q_3,q_1+q_4)$
&$(X_1,X_1+X_2,X_1+X_2+X_3)$
&$I_{1,2,3}^{(4)}$\\
&&&$(X_1,X_3+X_4,X_4)$
&$I_{1;4,3}^{(4)}$\\

\SetCell[r=2]{c}$R^1$
&\SetCell[r=2]{c}$(\vec k_2,\vec k_3,\vec k_4,\vec k_1)$
&\SetCell[r=2]{c}$(q_2+q_3,q_2+q_4,q_1+q_2)$
&$(X_2,X_2+X_3,X_2+X_3+X_4)$
&$I_{2,3,4}^{(4)}$\\
&&&$(X_2,X_1+X_4,X_1)$
&$I_{2;1,4}^{(4)}$\\

\SetCell[r=2]{c}$R^2$
&\SetCell[r=2]{c}$(\vec k_3,\vec k_4,\vec k_1,\vec k_2)$
&\SetCell[r=2]{c}$(q_3+q_4,q_1+q_3,q_2+q_3)$
&$(X_3,X_3+X_4,X_1+X_3+X_4)$
&$I_{3,4,1}^{(4)}$\\
&&&$(X_3,X_1+X_2,X_2)$
&$I_{3;2,1}^{(4)}$\\

\SetCell[r=2]{c}$R^3$
&\SetCell[r=2]{c}$(\vec k_4,\vec k_1,\vec k_2,\vec k_3)$
&\SetCell[r=2]{c}$(q_1+q_4,q_2+q_4,q_3+q_4)$
&$(X_4,X_1+X_4,X_1+X_2+X_4)$
&$I_{4,1,2}^{(4)}$\\
&&&$(X_4,X_2+X_3,X_3)$
&$I_{4;3,2}^{(4)}$\\

\SetCell[r=2]{c}$F_1$
&\SetCell[r=2]{c}$(\vec k_4,\vec k_3,\vec k_2,\vec k_1)$
&\SetCell[r=2]{c}$(q_1+q_4,q_1+q_3,q_1+q_2)$
&$(X_4,X_3+X_4,X_2+X_3+X_4)$
&$I_{4,3,2}^{(4)}$\\
&&&$(X_4,X_1+X_2,X_1)$
&$I_{4;1,2}^{(4)}$\\

\SetCell[r=2]{c}$F_2$
&\SetCell[r=2]{c}$(\vec k_1,\vec k_4,\vec k_3,\vec k_2)$
&\SetCell[r=2]{c}$(q_1+q_2,q_2+q_4,q_2+q_3)$
&$(X_1,X_1+X_4,X_1+X_3+X_4)$
&$I_{1,4,3}^{(4)}$\\
&&&$(X_1,X_2+X_3,X_2)$
&$I_{1;2,3}^{(4)}$\\

\SetCell[r=2]{c}$F_3$
&\SetCell[r=2]{c}$(\vec k_2,\vec k_1,\vec k_4,\vec k_3)$
&\SetCell[r=2]{c}$(q_2+q_3,q_1+q_3,q_3+q_4)$
&$(X_2,X_1+X_2,X_1+X_2+X_4)$
&$I_{2,1,4}^{(4)}$\\
&&&$(X_2,X_3+X_4,X_3)$
&$I_{2;3,4}^{(4)}$\\

\SetCell[r=2]{c}$F_4$
&\SetCell[r=2]{c}$(\vec k_3,\vec k_2,\vec k_1,\vec k_4)$
&\SetCell[r=2]{c}$(q_3+q_4,q_2+q_4,q_1+q_4)$
&$(X_3,X_2+X_3,X_1+X_2+X_3)$
&$I_{3,2,1}^{(4)}$\\
&&&$(X_3,X_1+X_4,X_4)$
&$I_{3;4,1}^{(4)}$\\
\hline[1pt]
\end{longtblr}
\endgroup

\noindent\textbf{Routing class 2.}

\begingroup\scriptsize\SetTblrTemplate{head}{empty}
\begin{longtblr}{ccccc}
\hline[1pt]
$D_4$ & $(\vec k_1',\vec k_2',\vec k_3',\vec k_4')$ & Edge Energies & Vertex Energies & Integrals\\
\hline
\SetCell[r=2]{c}$\mathbf 1$
&\SetCell[r=2]{c}$(\vec k_1,\vec k_2,\vec k_3,\vec k_4)$
&\SetCell[r=2]{c}$(q_2+q_3,q_1+q_3,q_1+q_4)$
&$(X_2,X_1+X_2,X_1+X_2+X_3)$
&$I_{2,1,3}^{(4)}$\\
&&&$(X_2,X_3+X_4,X_4)$
&$I_{2;4,3}^{(4)}$\\

\SetCell[r=2]{c}$R^1$
&\SetCell[r=2]{c}$(\vec k_2,\vec k_3,\vec k_4,\vec k_1)$
&\SetCell[r=2]{c}$(q_3+q_4,q_2+q_4,q_1+q_2)$
&$(X_3,X_2+X_3,X_2+X_3+X_4)$
&$I_{3,2,4}^{(4)}$\\
&&&$(X_3,X_1+X_4,X_1)$
&$I_{3;1,4}^{(4)}$\\

\SetCell[r=2]{c}$R^2$
&\SetCell[r=2]{c}$(\vec k_3,\vec k_4,\vec k_1,\vec k_2)$
&\SetCell[r=2]{c}$(q_1+q_4,q_1+q_3,q_2+q_3)$
&$(X_4,X_3+X_4,X_1+X_3+X_4)$
&$I_{4,3,1}^{(4)}$\\
&&&$(X_4,X_1+X_2,X_2)$
&$I_{4;2,1}^{(4)}$\\

\SetCell[r=2]{c}$R^3$
&\SetCell[r=2]{c}$(\vec k_4,\vec k_1,\vec k_2,\vec k_3)$
&\SetCell[r=2]{c}$(q_1+q_2,q_2+q_4,q_3+q_4)$
&$(X_1,X_1+X_4,X_1+X_2+X_4)$
&$I_{1,4,2}^{(4)}$\\
&&&$(X_1,X_2+X_3,X_3)$
&$I_{1;3,2}^{(4)}$\\

\SetCell[r=2]{c}$F_1$
&\SetCell[r=2]{c}$(\vec k_4,\vec k_3,\vec k_2,\vec k_1)$
&\SetCell[r=2]{c}$(q_3+q_4,q_1+q_3,q_1+q_2)$
&$(X_3,X_3+X_4,X_2+X_3+X_4)$
&$I_{3,4,2}^{(4)}$\\
&&&$(X_3,X_1+X_2,X_1)$
&$I_{3;1,2}^{(4)}$\\

\SetCell[r=2]{c}$F_2$
&\SetCell[r=2]{c}$(\vec k_1,\vec k_4,\vec k_3,\vec k_2)$
&\SetCell[r=2]{c}$(q_1+q_4,q_2+q_4,q_2+q_3)$
&$(X_4,X_1+X_4,X_1+X_3+X_4)$
&$I_{4,1,3}^{(4)}$\\
&&&$(X_4,X_2+X_3,X_2)$
&$I_{4;2,3}^{(4)}$\\

\SetCell[r=2]{c}$F_3$
&\SetCell[r=2]{c}$(\vec k_2,\vec k_1,\vec k_4,\vec k_3)$
&\SetCell[r=2]{c}$(q_1+q_2,q_1+q_3,q_3+q_4)$
&$(X_1,X_1+X_2,X_1+X_2+X_4)$
&$I_{1,2,4}^{(4)}$\\
&&&$(X_1,X_3+X_4,X_3)$
&$I_{1;3,4}^{(4)}$\\

\SetCell[r=2]{c}$F_4$
&\SetCell[r=2]{c}$(\vec k_3,\vec k_2,\vec k_1,\vec k_4)$
&\SetCell[r=2]{c}$(q_2+q_3,q_2+q_4,q_1+q_4)$
&$(X_2,X_2+X_3,X_1+X_2+X_3)$
&$I_{2,3,1}^{(4)}$\\
&&&$(X_2,X_1+X_4,X_4)$
&$I_{2;4,1}^{(4)}$\\
\hline[1pt]
\end{longtblr}
\endgroup

\noindent\textbf{Routing class 3.}

\begingroup\scriptsize\SetTblrTemplate{head}{empty}
\begin{longtblr}{ccccc}
\hline[1pt]
$D_4$ & $(\vec k_1',\vec k_2',\vec k_3',\vec k_4')$ & Edge Energies & Vertex Energies & Integrals\\
\hline
\SetCell[r=2]{c}$\mathbf 1$
&\SetCell[r=2]{c}$(\vec k_1,\vec k_2,\vec k_3,\vec k_4)$
&\SetCell[r=2]{c}$(q_1+q_2,q_1+q_2+q_3+q_4,q_1+q_4)$
&$(X_1,X_1+X_3,X_1+X_2+X_3)$
&$I_{1,3,2}^{(4)}$\\
&&&$(X_1,X_2+X_4,X_4)$
&$I_{1;4,2}^{(4)}$\\

\SetCell[r=2]{c}$R^1$
&\SetCell[r=2]{c}$(\vec k_2,\vec k_3,\vec k_4,\vec k_1)$
&\SetCell[r=2]{c}$(q_2+q_3,q_1+q_2+q_3+q_4,q_1+q_2)$
&$(X_2,X_2+X_4,X_2+X_3+X_4)$
&$I_{2,4,3}^{(4)}$\\
&&&$(X_2,X_1+X_3,X_1)$
&$I_{2;1,3}^{(4)}$\\

\SetCell[r=2]{c}$R^2$
&\SetCell[r=2]{c}$(\vec k_3,\vec k_4,\vec k_1,\vec k_2)$
&\SetCell[r=2]{c}$(q_3+q_4,q_1+q_2+q_3+q_4,q_2+q_3)$
&$(X_3,X_1+X_3,X_1+X_3+X_4)$
&$I_{3,1,4}^{(4)}$\\
&&&$(X_3,X_2+X_4,X_2)$
&$I_{3;2,4}^{(4)}$\\

\SetCell[r=2]{c}$R^3$
&\SetCell[r=2]{c}$(\vec k_4,\vec k_1,\vec k_2,\vec k_3)$
&\SetCell[r=2]{c}$(q_1+q_4,q_1+q_2+q_3+q_4,q_3+q_4)$
&$(X_4,X_2+X_4,X_1+X_2+X_4)$
&$I_{4,2,1}^{(4)}$\\
&&&$(X_4,X_1+X_3,X_3)$
&$I_{4;3,1}^{(4)}$\\

\SetCell[r=2]{c}$F_1$
&\SetCell[r=2]{c}$(\vec k_4,\vec k_3,\vec k_2,\vec k_1)$
&\SetCell[r=2]{c}$(q_1+q_4,q_1+q_2+q_3+q_4,q_1+q_2)$
&$(X_4,X_2+X_4,X_2+X_3+X_4)$
&$I_{4,2,3}^{(4)}$\\
&&&$(X_4,X_1+X_3,X_1)$
&$I_{4;1,3}^{(4)}$\\

\SetCell[r=2]{c}$F_2$
&\SetCell[r=2]{c}$(\vec k_1,\vec k_4,\vec k_3,\vec k_2)$
&\SetCell[r=2]{c}$(q_1+q_2,q_1+q_2+q_3+q_4,q_2+q_3)$
&$(X_1,X_1+X_3,X_1+X_3+X_4)$
&$I_{1,3,4}^{(4)}$\\
&&&$(X_1,X_2+X_4,X_2)$
&$I_{1;2,4}^{(4)}$\\

\SetCell[r=2]{c}$F_3$
&\SetCell[r=2]{c}$(\vec k_2,\vec k_1,\vec k_4,\vec k_3)$
&\SetCell[r=2]{c}$(q_2+q_3,q_1+q_2+q_3+q_4,q_3+q_4)$
&$(X_2,X_2+X_4,X_1+X_2+X_4)$
&$I_{2,4,1}^{(4)}$\\
&&&$(X_2,X_1+X_3,X_3)$
&$I_{2;3,1}^{(4)}$\\

\SetCell[r=2]{c}$F_4$
&\SetCell[r=2]{c}$(\vec k_3,\vec k_2,\vec k_1,\vec k_4)$
&\SetCell[r=2]{c}$(q_3+q_4,q_1+q_2+q_3+q_4,q_1+q_4)$
&$(X_3,X_1+X_3,X_1+X_2+X_3)$
&$I_{3,1,2}^{(4)}$\\
&&&$(X_3,X_2+X_4,X_4)$
&$I_{3;4,2}^{(4)}$\\
\hline[1pt]
\end{longtblr}
\endgroup

\subsection{Counting routing classes}
For any $n$, the number of the independent integrals under $D_n$ can be generally derived. We consider the following order of the vertex labeling
\begin{equation}
    \pi_1,\pi_2,\dots,\pi_n,
\end{equation}
and there are in total $n!$ of them. We further denote the edge energies corresponding to the tubes generated by these vertices:
\begin{equation}
    E(\pi_1),E(\pi_1,\pi_2),\dots,E(\pi_1,\dots,\pi_n).
\end{equation}
Note that this is equivalent to do the tubing in a reversal way
\begin{equation}
    E(\pi_1,\dots,\pi_n),\dots,E(\pi_{n-1},\pi_n),E(\pi_n),
\end{equation}
thus the only non-equivalent number of edge energies is $\frac{n!}{2}$, and from now on, we shall denote
\begin{equation}
    \Pi=S_n/(\pi\sim\pi^{\rm reverse}).
\end{equation}
The dihedral group $D_n$ acts on $\Pi$ by relabeling the vertices. By Burnside's lemma, the number of orbits is
\begin{equation}
    N=\frac{1}{|G|}\sum_{g\in G}|\operatorname{Fix}(g)|,
\end{equation}
where 
\begin{equation}
    \operatorname{Fix}(g)=\{x\in\Pi\mid gx=x\}.
\end{equation}
This gives only two possibilities
\begin{equation}
    g\pi=\pi\qquad\text{and}\qquad g\pi=\pi^{\rm reverse}.
\end{equation}
For the first case, the only way to make that is to let $g=\mathbf 1$, so
\begin{equation}
    |\operatorname{Fix}(\mathbf1)|=\frac{n!}{2}.
\end{equation}
For the second case, we need to seperate our discussion into parity of $n$:
\begin{enumerate}
    \item For $n=2m+1$, the only way achieve the second condition is to let $g=F_j$. And for each $F_j$ there are $m$ pairs with number of ordering $m!$ and for each pair there are 2 ways for arranging it in the front or in the back. After moduling out the $\mathbb Z_2$ symmetry, the total number is 
    \begin{equation}
       |\operatorname{Fix}(F_1)|+\cdots+|\operatorname{Fix}(F_{2m+1})|=(2m+1)2^{m-1}m!.
    \end{equation}
    \item For $n=2m$, we could choose either $R^{m}$ rotation or $m$ reflection on edges. These give us
    \begin{equation}
        |\operatorname{Fix}(F^{\rm edge}_1)|+\cdots+|\operatorname{Fix}(F^{\rm edge}_m)|+|\operatorname{Fix}(R^{m})|=(m+1)2^{m-1}m!.
    \end{equation}
\end{enumerate}
This gives us a total result of 
\begin{equation}\label{eq:independent_integral}
    {N_n=\begin{cases}\dfrac{1}{2(2m+1)}\left[\dfrac{(2m+1)!}{2}+(2m+1)2^{m-1}m! \right]&n=2m+1,\\
        \\
        \dfrac{1}{4m}\left[\dfrac{(2m)!}{2}+(m+1)2^{m-1}m!\right]&n=2m.
    \end{cases}}
\end{equation}
As some particular examples
\begin{equation}
    N_2=1,\quad N_3=1,\quad N_4=3,\quad N_5=8,\quad N_6=38,\dots
\end{equation}
We note that the number independent basic integrals under $D_{n}$ is dramatically less than the number of terms appeared either in \eqref{eq:time_integral} or \eqref{eq:reduction}, whose exact counting can be found in the Table \ref{tab:counts} below.
\begin{table}[htbp]
 \centering
 \begin{tabular}{@{}lccc@{}}
 \toprule
$n$ & Paired time orderings & Basic-integral terms
& Routing classes\\
 \midrule
 $2$ & $3$ & $2$ & $1$\\
 $3$ & $12$ & $9$ & $1$\\
 $4$ & $60$ & $48$ & $3$\\
 $n$ & $(n+1)!/2$ & $n\,n!/2$ & $N_n$\\
 \bottomrule
 \end{tabular}
 \caption{Numbers of time-ordering terms, basic-integral terms and inequivalent dihedral routing classes.}
 \label{tab:counts}
\end{table}

\section{Soft limits, collapsed channels and ultraviolet behavior}\label{app:limits}
\subsection{Soft limits of vertex energies}
We consider the soft limits of the vertex energies $X_i$ in the full
correlator $\mathcal I_n$, as analytic limits in the independent vertex energies, at fixed external spatial momenta. The time identities hold first at fixed loop distances.

If only $X_n$ is nonzero, the integrand is invariant under a common
translation of all $t_i$, except for $e^{-X_n|t_n|}$. Using
\begin{equation}
 \frac{e^{-q|t|}}{2q}
 =\int_{-\infty}^{\infty}\frac{\mathrm d\omega}{2\pi}
 \frac{e^{i\omega t}}{\omega^2+q^2},
\end{equation}
and integrating $t_n$ and the remaining relative times, we obtain the
exact identity
\begin{equation}\label{app:soft_one_hard}
 \left.\mathcal I_n\right|_{X_1=\cdots=X_{n-1}=0}
 =\frac{2}{X_n}\int\mathrm d^d\vec q_1
 \int_{-\infty}^{\infty}\frac{\mathrm d\omega}{2\pi}
 \prod_{i=1}^{n}\frac{1}{\omega^2+q_i^2}.
\end{equation}
In particular, this applies to the box with
$X_1=X_2=X_3=0$, $X_4>0$.

For the box with two nonvanishing vertex energies, we can also set
$X_1=X_3=0$ while keeping $X_2,X_4>0$ and the $q_i$ fixed.
Integrating $t_1,t_3$ first gives the two convolution kernels
\begin{equation}
 \int_{-\infty}^{\infty}\mathrm dt_1\,
 \frac{e^{-q_1|t_1-t_4|}}{2q_1}
 \frac{e^{-q_2|t_2-t_1|}}{2q_2}
 =\frac{\frac{e^{-q_1|t_2-t_4|}}{2q_1}
 -\frac{e^{-q_2|t_2-t_4|}}{2q_2}}{q_2^2-q_1^2},
\end{equation}
with the analogous expression for $q_3,q_4$. Combining this with
\begin{equation}
 \int_{-\infty}^{\infty}\mathrm dt_4\,
 e^{-X_2|t_4+\tau|-X_4|t_4|}
 =\frac{2\bigl(X_2e^{-X_4\tau}-X_4e^{-X_2\tau}\bigr)}
 {X_2^2-X_4^2},
 \qquad \tau\geq 0,
\end{equation}
we find
\begin{align}\label{app:soft_two_hard}
 \left.\mathcal I_4\right|_{X_1=X_3=0}
 ={}&4\int\mathrm d^d\vec q_1\int_0^\infty\mathrm d\tau\,
 \frac{X_2e^{-X_4\tau}-X_4e^{-X_2\tau}}{X_2^2-X_4^2}
\frac{\frac{e^{-q_1\tau}}{2q_1}-\frac{e^{-q_2\tau}}{2q_2}}
 {q_2^2-q_1^2}
 \frac{\frac{e^{-q_3\tau}}{2q_3}-\frac{e^{-q_4\tau}}{2q_4}}
 {q_4^2-q_3^2}.
\end{align}

\subsection{Collapsed-channel coefficient of the box}
We consider the collapsed channel $\vec K_2=\vec k_1+\vec k_2\to0$
with $X_i>0$ and $|\vec k_1|,|\vec k_3|$ fixed and generic.
The nonanalytic behavior of inflationary correlators in collapsed
momentum channels, and its relation to factorization and the OPE,
has been studied in~\cite{Qin:2023bjk,Xianyu:2022jwk}.
Here we derive the leading nonanalytic coefficient of the box
directly from its Baikov representation and the full-time integral. For
$n=4$, all basic tubing integrals share the measure
\begin{equation}
 \frac{\mathrm dq_1\cdots\mathrm dq_4}{(\det H_4)^{(d-4)/2}}
 \left[\det\operatorname{Gram}(\vec q_1,\vec K_1,\vec K_2,\vec K_3)
 \right]^{(d-5)/2},
\end{equation}
up to the common normalization and their rational tubing denominators.
Since $q_3=|\vec q_1+\vec K_2|$, the Baikov domain satisfies
\begin{equation}
 |q_1-q_3|\leq |\vec K_2|\leq q_1+q_3.
\end{equation}
The leading nonanalytic contribution comes from the region
\begin{equation}
 q_1,q_3=O(|\vec K_2|),\qquad
 q_2=|\vec k_1|+O(|\vec K_2|),\qquad
 q_4=|\vec k_3|+O(|\vec K_2|).
\end{equation}
For a generic direction of $\vec K_2$, the Gram determinants and
integration measure scale as
\begin{equation}
 \det H_4=O(|\vec K_2|^2),\quad
 \det\operatorname{Gram}(\vec q_1,\vec K_1,\vec K_2,\vec K_3)
 =O(|\vec K_2|^4),\quad
 \mathrm dq_1\cdots\mathrm dq_4=O(|\vec K_2|^4).
\end{equation}
Thus the soft region of the Baikov measure scales as
$|\vec K_2|^{4-(d-4)+2(d-5)}=|\vec K_2|^{d-2}$.

To obtain the coefficient for the full correlator, we must
recombine the tubing terms. Equivalently, in its full-time
representation the two soft propagators decouple the pairs $(1,2)$
and $(3,4)$. The hard-pair time integrals give
\begin{align}\label{app:ope_hard_pair}
 &\int_{-\infty}^{\infty}\mathrm dt_i\,\mathrm dt_{i+1}\,
 \frac{e^{-X_i|t_i|-X_{i+1}|t_{i+1}|-
 |\vec k_i||t_i-t_{i+1}|}}{2|\vec k_i|}=\frac{2(X_i+X_{i+1}+|\vec k_i|)}
 {|\vec k_i|(X_i+X_{i+1})(X_i+|\vec k_i|)
 (X_{i+1}+|\vec k_i|)},\quad i=1,3.
\end{align}
Consequently
\begin{align}\label{app:ope_factorization}
 \left.\mathcal I_4\right|_{\rm nonanalytic}
 ={}&\frac14
 \prod_{i=1,3}
 \frac{2(X_i+X_{i+1}+|\vec k_i|)}
 {|\vec k_i|(X_i+X_{i+1})(X_i+|\vec k_i|)
 (X_{i+1}+|\vec k_i|)}
\left.\int\frac{\mathrm d^d\vec q_1}
 {|\vec q_1|\,|\vec q_1+\vec K_2|}
 \right|_{\rm nonanalytic}.
\end{align}

Using the Schwinger representation
\begin{equation}
 \frac1{|\vec q_1|\,|\vec q_1+\vec K_2|}
 =\frac1\pi\int_0^\infty\frac{\mathrm d\alpha\,\mathrm d\beta}
 {\sqrt{\alpha\beta}}
 e^{-\alpha\vec q_1^{\,2}-\beta(\vec q_1+\vec K_2)^2},
\end{equation}
the Gaussian integral and the remaining beta integral give
\begin{equation}\label{app:ope_soft_integral}
 \int\frac{\mathrm d^d\vec q_1}
 {|\vec q_1|\,|\vec q_1+\vec K_2|}
 =\pi^{\frac d2-1}
 \frac{\Gamma\bigl(\frac{d-1}{2}\bigr)^2
 \Gamma\bigl(1-\frac d2\bigr)}{\Gamma(d-1)}
 (\vec K_2^{\,2})^{\frac d2-1}.
\end{equation}
After subtracting analytic terms in $d=4-2\epsilon$, we obtain
\begin{align}
 \left.\mathcal I_4\right|_{\substack{d=4\\\rm leading\,nonanalytic}}
 & =\frac{\pi^2}{32}
 \prod_{i=1,3}
 \frac{2(X_i+X_{i+1}+|\vec k_i|)}
 {|\vec k_i|(X_i+X_{i+1})(X_i+|\vec k_i|)
 (X_{i+1}+|\vec k_i|)}
 \vec K_2^{\,2}\log\frac{\vec K_2^{\,2}}{\mu^2},
 \\
 \left.\mathcal I_4\right|_{\substack{d=5\\\rm leading\,nonanalytic}}
 & =\frac{\pi^2}{18}
 \prod_{i=1,3}
 \frac{2(X_i+X_{i+1}+|\vec k_i|)}
 {|\vec k_i|(X_i+X_{i+1})(X_i+|\vec k_i|)
 (X_{i+1}+|\vec k_i|)}
 |\vec K_2|^3.
\end{align}
These are leading nonanalytic terms. Analytic matching terms and higher-order nonanalytic contributions are not included. The soft integral is defined by dimensional continuation; the UV pole of this region is matched to analytic terms from the hard region.

\subsection{Ultraviolet divergence at \texorpdfstring{$d=2n-1$}{d=2n-1}}
The asymptotic behavior of cosmological integrals and the
separation of momentum regions have been studied in
\cite{Benincasa:2024lxe,Beneke:2023wmt}.
Here we show that, at $d=2n-1$, the kernel UV coefficient
is $2/X_{\rm tot}$ times that of the associated flat-space
integral with frequency measure $\dd\omega/(2\pi)$.

To see this, use the full-time representation of $\mathcal I_n$ and
take $|\vec q_1|\to\infty$ with fixed $X_i$ and $\vec k_i$. Then
$q_i=|\vec q_1|+O(1)$ for every $i$, and the relative times
$t_i-t_j$ localize to $O(|\vec q_1|^{-1})$. To leading order
\begin{equation}
 \prod_{i=1}^{n}e^{-X_i|t_i|}
 \longrightarrow e^{-\left(\sum_{i=1}^{n}X_i\right)|t_n|},
 \qquad
 \int_{-\infty}^{\infty}\mathrm dt_n\,
 e^{-\left(\sum_{i=1}^{n}X_i\right)|t_n|}
 =\frac{2}{\sum_{i=1}^{n}X_i}.
\end{equation}
The remaining $(n-1)$ relative-time integrals coincide with the
zero-external-frequency flat-space loop energy integral. Thus, at
leading UV order
\begin{align}
 \mathcal I_n\big|_{\rm UV}
 &\sim\frac{2}{\sum_{i=1}^{n}X_i}
 \int\mathrm d^d\vec q_1
 \int_{-\infty}^{\infty}\frac{\mathrm d\omega}{2\pi}
 \frac{1}{(\omega^2+\vec q_1^{\,2})^n}
=\frac{\Gamma\bigl(n-\frac12\bigr)}
 {\sqrt\pi\,\Gamma(n)\sum_{i=1}^{n}X_i}
 \int\frac{\mathrm d^d\vec q_1}{|\vec q_1|^{2n-1}}.
\end{align}
At $d=2n-1$, the radial integral is logarithmic, giving
\begin{equation}\label{app:uv_ngon}
\left.\mathcal I_n\right|_{d=2n-1,\,\rm UV}
 =\frac{2\pi^{n-1}}{\Gamma(n)\sum_{i=1}^{n}X_i}
 \log\frac{\Lambda}{\mu}
 +\text{UV-finite terms}.
\end{equation}
In precisely the same normalization for the flat-space loop-energy
measure, the standard Euclidean $(d+1)$-dimensional $n$-gon obeys
\begin{equation}
 \left.\int\mathrm d^d\vec q_1
 \int_{-\infty}^{\infty}\frac{\mathrm d\omega}{2\pi}
 \prod_{i=1}^{n}\frac{1}{\omega^2+q_i^2}
 \right|_{d=2n-1,\,\rm UV}
 =\frac{\pi^{n-1}}{\Gamma(n)}\log\frac{\Lambda}{\mu}
 +\text{UV-finite terms}.
\end{equation}
Hence the logarithmic UV coefficients agree after extracting the
universal full-time factor $2/(\sum_i X_i)$.

\bibliographystyle{JHEP}
\bibliography{ngon_refs}
\end{document}